\documentclass[a4paper,11pt]{article}

\usepackage{titlesec}
\usepackage[titletoc,toc,title]{appendix}

\usepackage{graphicx,epstopdf}
\usepackage{dsfont, color, braket}
\usepackage{braket}
\usepackage[space]{grffile}

\usepackage[T1]{fontenc}
\usepackage[utf8]{inputenc}
\usepackage{graphicx,tabularx}

\usepackage{amsmath,amssymb,amsthm,float,mathrsfs}
\usepackage[affil-it]{authblk}
\usepackage{dsfont}

\def\beq{\begin{equation}}
\def\eeq{\end{equation}}
\def\beqa{\begin{eqnarray}}
\def\eeqa{\end{eqnarray}}

\newcommand{\eq}[1]{Eq.(#1)}
\newcommand{\fig}[1]{Fig. #1}

\numberwithin{equation}{section}

\title{Non-equilibrium dynamics of non-linear Jaynes-Cummings model in cavity arrays}
\author[]{Ji\v{r}\'{i} Min\'{a}\v{r}, \c{S}ebnem G\"{u}ne\c{s} S\"{o}yler and Igor Lesanovsky}
\affil[1]{School of Physics and Astronomy, University of Nottingham, University Park, NG7 2RD, Nottigham, United Kingdom}
\date{\vspace{-1ex}} 

\begin{document}

\maketitle

\begin{abstract}
We analyze in detail an open cavity array using mean-field description, where each cavity field is coupled to a number of three-level atoms. Such system is highly tunable and can be described by a Jaynes-Cummings like Hamiltonian with additional non-linear terms. In the single cavity case we provide simple analytic solutions and show, that the system features a bistable region. The extra non-linear term gives rise to a rich dynamical behaviour including occurrence of limit cycles through Hopf bifurcations. In the limit of large non-linearity, the system exhibits an Ising like phase transition as the coupling between light and matter is varied. We then discuss how these results extend to the two-dimensional case.
\end{abstract}

\section{Introduction}

The use of cavity QED tools is now ubiquitous across different areas of physics ranging from quantum information \cite[p. 75]{Everitt_2005} to detection of dark matter \cite{Bradley_2003}. Specifically, the atoms held in optical cavities play a vital role in studies of many-body physics \cite{Maschler_2008}. Such systems are natural implementations of many-body Jaynes-Cummings (JC) or Dicke Hamiltonians \cite{Dimer_2007}. Their high tunability and the possibility of achieving strong light matter coupling or probing the dynamics in real time make them very attractive experimental platforms. The prospect of probing phase transitions and the associated critical phenomena with these platforms have been put forward e.g. in \cite{Gammelmark_2011,Bakemeier_2012,Castanos_2012} and the non-equilibrium dynamics of the Dicke model has been theoretically investigated in \cite{Bhaseen_2012}. Specific many-body phenomena that can be studied with cavity QED include for example the physics of spin glasses \cite{Strack_2011, Buchhold_2013} or the self-organization of the atomic motion \cite{Domokos_2002,Zippilli_2004,Asboth_2005}, to name a few. The self-organization has been subsequently observed in the experiments \cite{Black_2003,Baumann_2010}.

So far we have mentioned only studies concerning a single cavity - generalizations to multiple cavity arrays implementing the Hubbard physics have been reviewed in \cite{Hartmann_2008a}. Although appealing in principle, the realization of many efficiently coupled cavities, each hosting a discrete-level quantum system is a challenging task. To make such experiment scalable requires miniaturization of the cavities. One possibility is the use of microcavities in photonic crystals \cite{Vuckovic_2001,Vuckovic_2003}. A further option is to use integrated optical circuits, where in principle arbitrary waveguide forms can be created with high precision by laser engraving in the silica substrate \cite{Marshall_2009}. They have been successfully used for the demonstration of a quantum gate operation \cite{Crespi_2011}, creation of classical and quantum correlations \cite{Bromberg_2009,Keil_2011}, multi-photon entangled state preparation \cite{Grafe_2014}, quantum random walk \cite{Peruzzo_2010}, discrete Fourier transform \cite{Weimann_2015} or Bloch oscillations \cite{Lebugle_2015}.

While it has been demonstrated that cavities can be fabricated by creating the Bragg grating during the laser writing process \cite{Marshall_2006, Thiel_2015}, there is now an active experimental effort to combine the waveguides with atomic microtraps on a single device \cite{Derntl_2014, Potts_2016}.

Motivated by these developments and the prospects of studying many-body physics using integrated optical circuits with trapped atoms, we theoretically analyze the non-equilibrium physics of such system, which we take to be a two-dimensional cavity array, where each cavity hosts a number of atoms. We first derive the effective Hamiltonian describing the system in Sec. \ref{sec:Model}. Using this Hamiltonian we find mean-field (MF) equations of motion whose solutions feature rich dynamics, including bistable behaviour, Ising like phase transition or the occurrence of limit cycles, which we discuss in detail in Sec. {\ref{sec:MF}}. We then conclude in Sec. \ref{sec:Conclusion}.

\begin{center}
	\begin{figure}[t!p]
  	\includegraphics[width=12cm]{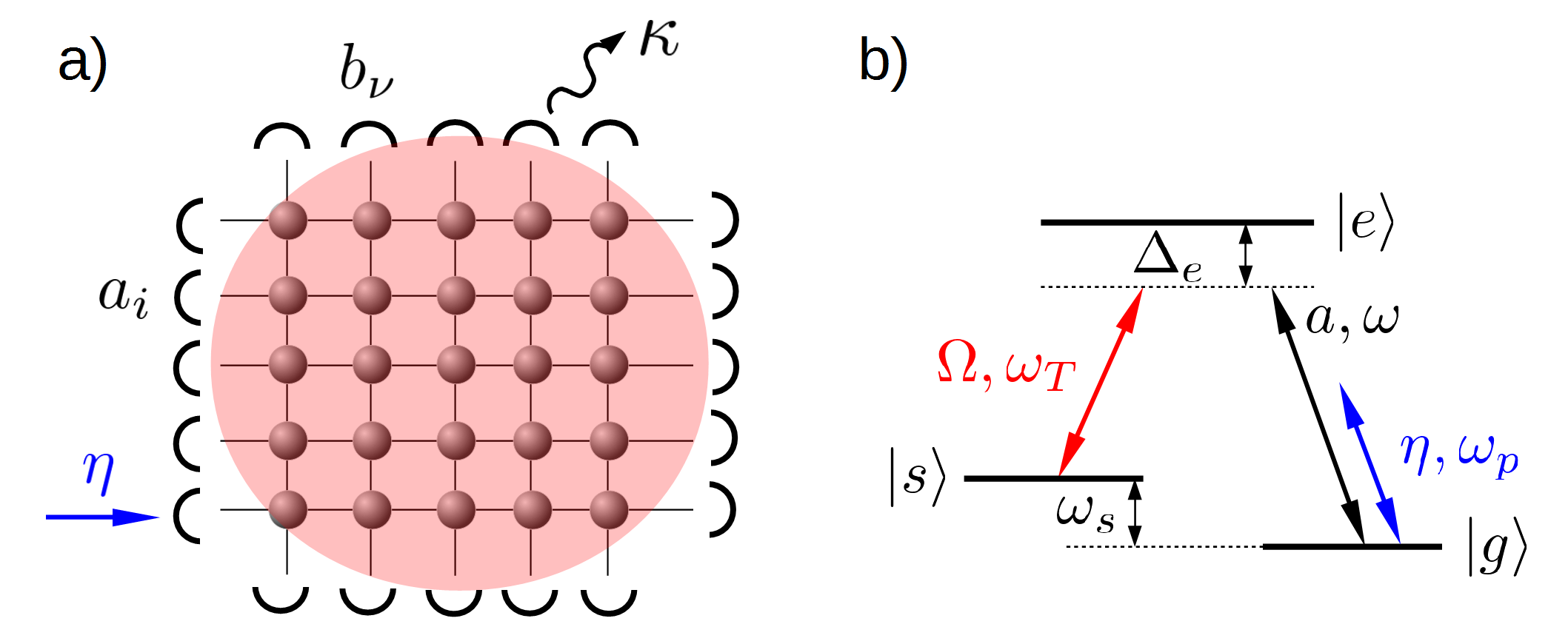} 
  	\caption{(Color online) (a) Scheme of the envisioned experimental setup. The atoms are located at the intersections of the cavity modes and are represented by the small spheres. The classical light field $\Omega$ is sketched by the large red circle. (b) Atomic level scheme (only a single $a$ mode is shown, see text for details).}
  	\label{fig:Scheme}
 	\end{figure}
\end{center}

\section{The Model}
\label{sec:Model}

We consider a two-dimensional array of identical three-level atoms each coupled to a horizontal cavity mode $a_i$ and vertical cavity mode $b_\nu$ with the same coupling constant $g$, \fig{\ref{fig:Scheme}a}. From here on we use latin and greek indices to denote horizontal and vertical degrees of freedom respectively. In order to simplify the notation, we omit the sum symbols and summation over occurrences of indices labeling the \emph{spatial} position is understood (i.e. $o_j \equiv \sum_j o_j$). We will use indices with bar, e.g. $\bar{j}$, when we want to emphasize that the index $\bar{j}$ is fixed and is not summed over. The atomic level structure is depicted in \fig{\ref{fig:Scheme}b}. The $\ket{g}-\ket{e}$ transition is coupled to cavity mode $a_{\bar{i}}$ ($b_{\bar \nu}$) with frequency $\omega_{\bar i}$ ($\omega_{\bar \nu}$). The $\ket{s}-\ket{e}$ transition is coupled using a strong transverse classical light, whose direction of propagation is perpendicular to the plane of the cavity array and which has Rabi frequency $\Omega$, carrier frequency $\omega_T$ and detuning $\Delta_e$. 

The main reason for choosing a three level $\Lambda$ system is that it allows, in the limit of large $\Delta_e$, to adiabatically eliminate the excited state $\ket{e}$ in order to avoid the losses due to the spontaneous emission and to obtain an effective Hamiltonian in the ground state manifold subspace. This effective Hamiltonian features a high tunability, cf. below. We now proceed with the derivation of the effective Hamiltonian. We discuss the issue of cavity gain and loss later in Sec. \ref{sec:MF}.

The system is described by the Hamiltonian (in the rotating wave approximation)
\beqa
	H &=& \omega_i a_i^\dag a_i + \omega_\nu b_\nu^\dag b_\nu + \omega_e \ket{e}_{i\nu}\bra{e}_{i\nu} + \omega_s \ket{s}_{i\nu}\bra{s}_{i \nu} \nonumber \\
	&& + \Omega(\ket{e}_{i\nu}\bra{s}_{i\nu} {\rm e}^{-i \omega_T t} + \ket{s}_{i\nu}\bra{e}_{i\nu} {\rm e}^{i \omega_T t}) \nonumber \\
	&& + g (\ket{e}_{i\nu}\bra{g}_{i\nu} (a_i + b_\nu) + {\rm h.c.}).
	\label{eq:H RWA 2D}
\eeqa
The excited state $\ket{e}$ can be adiabatically eliminated in the standard way, see Appendix \ref{app:Ad.el} for details. The resulting effective Hamiltonian reads

\beq
	H = \Delta_i a_i^\dag a_i + \Delta_\nu b_\nu^\dag b_\nu + \frac{\tilde{\omega}_{a,i\nu}}{2} \sigma^z_{i\nu} + \tilde{g}\left( \sigma^+_{i\nu} (a_i + b_\nu) + {\rm h.c.} \right)+F_{i \nu},	
	\label{eq:H eff RWA 2D}
\eeq
where $\sigma$ are the usual Pauli matrices in the $\{ \ket{s},\ket{g} \}$ basis,
\beqa
	\tilde{\omega}_{a,\bar{i} \bar{\nu}} &=& \Delta_s - \frac{\Omega^2 - g^2 \left( a_{\bar i}^\dag + b_{\bar \nu}^\dag  \right) \left( a_{\bar i} + b_{\bar \nu}  \right) }{\Delta_e} \nonumber \\
	\tilde{g} &=& -\frac{g \Omega}{\Delta_e} \nonumber \\
	F_{\bar{i} \bar{\nu}} &=& \frac{1}{2} \left( \Delta_s - \frac{\Omega^2 + g^2 \left( a_{\bar i}^\dag + b_{\bar \nu}^\dag  \right) \left( a_{\bar i} + b_{\bar \nu}  \right) }{\Delta_e} \right).
	\label{eq:pars eff}
\eeqa
and $\Delta_x = \omega_x - \omega_{\rm aux}$, $x=\bar{i},\bar{\nu},e$ and $\Delta_s = \omega_s - (\omega_{\rm aux} - \omega_T)$. Here, $\omega_{\rm aux}$ is an auxiliary frequency used in the adiabatic elimination which, in principle, can be chosen arbitrarily. Its physical motivation and interpretation will be discussed momentarily in Sec. \ref{sec:Single cavity}.

\section{Mean-field treatment of the non-linear JC model}
\label{sec:MF}

The effective Hamiltonian (\ref{eq:H eff RWA 2D}) is the starting point in this section, where we analyze the effect of cavity loss and pump on the dynamics. Also, it can be seen from the form of the Hamiltonian and the expressions (\ref{eq:pars eff}), that the parameters are highly tunable through varying $\Omega, \Delta_e$ and $\Delta_s$.

We first perform a MF analysis of a simpler system with only one cavity mode, which already contains rich physics as we show below. We then turn back to the multi-mode two-dimensional setup in Sec. \ref{sec:2D}.

\subsection{Single cavity}
\label{sec:Single cavity}

\noindent \emph{Cavity without pump}

\noindent Lets first consider the Hamiltonian (\ref{eq:H eff RWA 2D}) in the single cavity limit with the single mode $a$, i.e. we drop the indices $i$ and $\nu$. We rewrite the Hamiltonian as
\beq
	H = \left( \frac{\Delta_{\rm at}}{2} + \lambda a^\dag a \right) \Sigma^z + \Delta_{\rm ph} a^\dag a + \tilde{g}\left(\Sigma^+ a + a^\dag \Sigma^- \right),
	\label{eq:H JC}
\eeq
where $\Sigma^{z,\pm} = \sigma^{z,\pm}_i$ are the global spin operators and 
\beqa
	\Delta_{\rm at} &=& \Delta_s - \frac{\Omega^2}{\Delta_e} \nonumber \\
	\Delta_{\rm ph} &=& \Delta - \frac{g^2}{2 \Delta_e} = \omega - \omega_{\rm aux} - \frac{g^2}{2 \Delta_e} \nonumber \\
	\lambda &=& -\frac{g^2}{2 \Delta_e},
\eeqa
with $\omega$ being the cavity frequency. Note that the model (\ref{eq:H JC}) without the $\lambda$ term is the usual JC model \cite{Jaynes_1963}. Below we show, that the $\lambda$ term is indeed at the origin of intriguing system dynamics (see also \cite{Vogel_1989, SebaweAbdalla_1990, Vogel_1995, Joshi_2000, Sivakumar_2000, Budini_2003, Sivakumar_2004, Singh_2011} for various other non-linear extensions of the JC model).

One can now derive the equation of motion for the operator $o$ according to $\dot{o} = -i [o,H]$, where $H$ is given by (\ref{eq:H JC}). At the same time, any realistic cavity is subject to a decay of the electromagnetic field into the environment. The dissipation process is typically described by means of a master equation (see e.g. \cite{Walls_1994}), which corresponds to an extra term in the equation of motion for the cavity mode operator, $\dot{a} \propto -\kappa a$, where $\kappa$ denotes the cavity loss rate (see \fig{\ref{fig:Scheme}a}). Introducing the expectation values of the operators $\alpha = \braket{a}, s = \braket{\Sigma^-}, w=\braket{\Sigma^z}$, it is now straightforward to derive the the MF equations of motion, which read

\begin{subequations}
\label{eq:EoMs MF}
\begin{align}
   i\dot{\alpha} &= \left(\lambda w + \Delta_{\rm ph}  - i \kappa \right)\alpha + \tilde{g} s \label{eq:EoMs MF a} \\
  i \dot{s} &= \left(\Delta_{\rm at} + 2 \lambda |\alpha|^2 \right) s - \tilde{g} w \alpha \label{eq:EoMs MF s} \\
  i \dot{w} &= 2 \tilde{g} \left( s^* \alpha - \alpha^* s \right). \label{eq:EoMs MF w}
\end{align}
\end{subequations}
In the derivation we have used the MF decoupling $\braket{a^\dag \Sigma^-} = \alpha^* w$ and $\braket{\Sigma^z a} = w \alpha$. We have also neglected the spin decay on the transition $\ket{g}-\ket{s}$\footnote{This is a justified assumption in implementations with real atoms as the levels $\ket{g},\ket{s}$ typically belong to some ground state manifold, where only magnetic dipole transitions are allowed between the states of that manifold. In turn, the spin decay is negligible compared to the cavity decay.}

First, we wish to find a steady state solution of the equations of motion (\ref{eq:EoMs MF}). From (\ref{eq:EoMs MF a}) and (\ref{eq:EoMs MF w}) we have
\beqa
  \alpha  &=& -\frac{\tilde{g}}{\lambda w + \Delta_{\rm ph}  - i \kappa} s \equiv C s \label{eq:cond1} \\
  s^* \alpha &=& \alpha^* s. \label{eq:cond2}
\eeqa
Substituting \eq{\ref{eq:cond1}} to \eq{\ref{eq:cond2}} yields
\beq
  C |s|^2 = C^* |s|^2.
  \label{eq:restr}
\eeq
When $C$ is complex, which occurs only for non-zero cavity decay $\kappa \neq 0$, the condition \eq{\ref{eq:restr}} can be satisfied if $s=0$ (and consequently $\alpha=0$), which results in a trivial solution with empty cavity and all spins down (up) in the steady state. This indicates that in order to obtain some non-trivial physics in the steady state limit and to counteract the cavity losses, one needs to input energy into the system\footnote{Note that this is in sharp contrast with the full Dicke Hamiltonian, where the presence of the counterrotating terms guarantees non-trivial solutions even in the absence of the pumping \cite{Dimer_2007,Bhaseen_2012}. In the model we study, the absence of the counterrotating terms is a direct consequence of applying the rotating wave approximation, as the cavity modes and the $\ket{g}$--$\ket{e}$ transition are taken to be at optical frequencies.}. In our case a natural choice is either through the pumping of the cavity mode or driving directly the $\ket{g}-\ket{s}$ transition. We focus here on the former case only.

~\\
\noindent \emph{Cavity with pump} 

\noindent The cavity pump can be described by the Hamiltonian
\beq
	H_{\rm pump} = \eta a^\dag {\rm e}^{-i \omega_p t} + \eta^* a {\rm e}^{i \omega_p t}.
\eeq
When adding $H_{\rm pump}$ to the system Hamiltonian (\ref{eq:H RWA 2D}), the explicit time dependence in the total Hamiltonian can be removed if the auxiliary frequency $\omega_{\rm aux}$ in (\ref{eq:pars eff}) is chosen such that it is the frequency of the cavity pump, $\omega_{\rm aux}=\omega_p$. The MF equations of motion (\ref{eq:EoMs MF}) then become
\beqa
  \dot{\alpha}_R &=& - \kappa \alpha_R + \left(\lambda w + \Delta_{\rm ph} \right) \alpha_I - \frac{\tilde{g}}{2} s_y + \eta_I \nonumber \\
  \dot{\alpha}_I &=& -\left(\lambda w + \Delta_{\rm ph} \right) \alpha_R - \kappa \alpha_I - \frac{\tilde{g}}{2} s_x  - \eta_R \nonumber \\
  \dot{s}_x &=& - 2 \tilde{g} w \alpha_I - \left(\Delta_{\rm at} + 2 \lambda |\alpha|^2 \right) s_y \nonumber \\
  \dot{s}_y &=& - 2 \tilde{g} w \alpha_R + \left(\Delta_{\rm at} + 2 \lambda |\alpha|^2 \right) s_x \nonumber \\ 
  \dot{w} &=& 2 \tilde{g} \left( \alpha_R s_y + \alpha_I s_x \right),
  \label{eq:EoMs MF pump real}
\eeqa
where we have introduced real variables through $\alpha=\alpha_R + i \alpha_I$, $s=1/2(s_x - i s_y)$ and $\eta = \eta_R + i \eta_I$.

Note that the equations (\ref{eq:EoMs MF pump real}) imply the conservation of the total spin
\beq
	w^2 + 4 |s|^2 = N^2,
	\label{eq:spin conservation}
\eeq
where $N$ is the total number of the spins. This can be easily verified as
\beq
  \partial_t S^2 = \partial_t (s_x^2 + s_y^2 + w^2) =  \partial_t (4 s s^* + w^2) = 0,
  \label{eq:diff spin conservation}
\eeq
where we have parametrized the total spin as $\vec{S} = (s_x,s_y,w)$.

\begin{center}
	\begin{figure}[t!p]	
	\hspace*{-1cm}
  	\includegraphics[width=13cm]{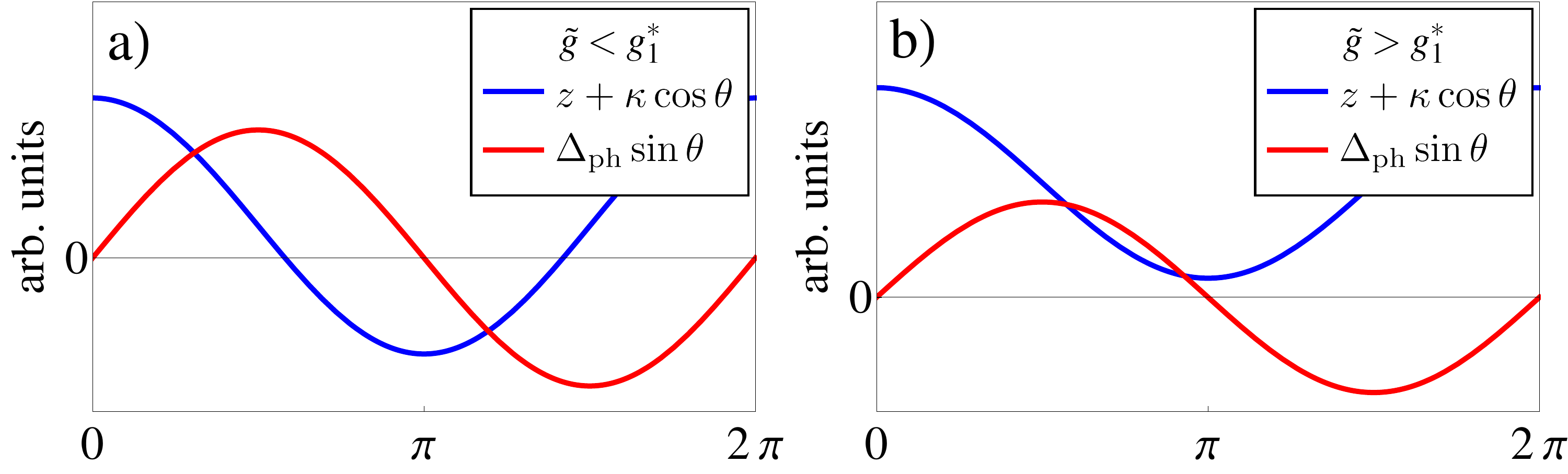} 
  	\caption{(Color online) Graphical representation of (\ref{eq:gon eq}) for (a) $\tilde{g} < g_1^*$ and (b) $\tilde{g} > g_1^*$. Left hand side (right hand side) of (\ref{eq:gon eq}) is represented by solid blue (red) lines. See text for details.}
  	\label{fig:gon}
 	\end{figure}
\end{center}
\subsubsection{$\lambda=0$ regime}
\label{sec:linear}

In order to investigate the steady state solutions, we first put $\lambda=0$ to further simplify the problem (see e.g. \cite{Koch_2009, Emary_2003,Bhaseen_2012} for related studies of the phase diagram of the JC and Dicke models). We then turn back to the situation with $\lambda \neq 0$ in the next section. With these simplifications, the real equations for the steady state read

\begin{subequations}
\label{eq:EoMs MF pump real SS}
\begin{align}
   0 &= - \kappa \alpha_R + \Delta_{\rm ph} \alpha_I - \frac{\tilde{g}}{2} s_y + \eta_I \label{eq:EoMs MF pump real SS aR} \\
   0 &= - \Delta_{\rm ph} \alpha_R - \kappa \alpha_I - \frac{\tilde{g}}{2} s_x  - \eta_R \label{eq:EoMs MF pump real SS aI} \\
   0 &= - 2 \tilde{g} w \alpha_I - \Delta_{\rm at} s_y \label{eq:EoMs MF pump real SS sx} \\
   0 &= - 2 \tilde{g} w \alpha_R + \Delta_{\rm at} s_x \label{eq:EoMs MF pump real SS sy} \\ 
   0 &= 2 \tilde{g} \left( \alpha_R s_y + \alpha_I s_x \right). \label{eq:EoMs MF pump real SS w}
\end{align}
\end{subequations}

Solving the set (\ref{eq:EoMs MF pump real SS aR},\ref{eq:EoMs MF pump real SS aI}) for $\alpha_R,\alpha_I$ and substituting to (\ref{eq:EoMs MF pump real SS sx}-\ref{eq:EoMs MF pump real SS w}) we obtain for the spin conservation (\ref{eq:spin conservation})
\beqa
  w^2 - N^2  = \frac{-4 \tilde{g}^2 w^2 \eta^2}{g^4 w^2 + \Delta_{\rm at}^2 \kappa^2 + \Delta_{\rm ph} \Delta_{\rm at} (2 g^2 w + \Delta_{\rm ph} \Delta_{\rm at})},
  \label{eq:poly w}
\eeqa
where $\eta^2 = \eta_R^2 + \eta_I^2$. This is a 4th order polynomial for $w$ and its solutions in terms of radicals can in principle be found yielding rather complicated expressions, which are not of much practical use. Instead, we will analyze the properties of (\ref{eq:poly w}) as follows. Since $w/N \in [-1; 1]$, the lhs of (\ref{eq:poly w}) is non-positive, namely $(w/N)^2-1 \in [-1,0]$. At the same time the nominator of the rhs is clearly non-positive, so that the non-positivity of the rhs requires the quadratic polynomial in the denominator to be non-negative. We thus have
\beq
  g^4 w^2 + \Delta_{\rm at}^2 \kappa^2 + \Delta_{\rm ph} \Delta_{\rm at} (2 g^2 w + \Delta_{\rm ph} \Delta_{\rm at}) \geq 0.
  \label{eq:cond poly}
\eeq
The roots of this polynomial read
\beq
  w_\pm = \frac{-2 \tilde{g}^2 \Delta_{\rm ph} \Delta_{\rm at} \pm \sqrt{-4 \tilde{g}^4 \Delta_{\rm at}^2 \kappa^2} }{2 \tilde{g}^4}.
\eeq
It can be immediately seen that for $\kappa \neq 0$, any $\Delta_{\rm at} \neq 0$ would yield imaginary $w$. This however does not mean that a solution of the original constraint (\ref{eq:poly w}) is not possible for both $\kappa, \Delta_{\rm at} \neq 0$. Rather, it means that the point $\Delta_{\rm at}=0$ has some particular properties which we will investigate further.

For $\Delta_{\rm at}=0$, the steady state equations read
\begin{subequations}
\label{eq:SS l0 D0}
\begin{align}
  \kappa \alpha_R - \Delta_{\rm ph} \alpha_I &= - \frac{\tilde{g}}{2} s_y + \eta_I \label{eq:SS l0 D0 aR} \\
  \Delta_{\rm ph} \alpha_R + \kappa \alpha_I &= - \frac{\tilde{g}}{2} s_x  - \eta_R \label{eq:SS l0 D0 aI} \\
  0 &= w \alpha_I \label{eq:SS l0 D0 sx} \\
  0 &= w \alpha_R \label{eq:SS l0 D0 sy} \\ 
  \alpha_R s_y &=  - \alpha_I s_x. \label{eq:SS l0 D0 w}
\end{align}
\end{subequations}
(\ref{eq:SS l0 D0 sx},\ref{eq:SS l0 D0 sy}) imply either $w=0$ or $\alpha_R = \alpha_I = 0$. If $\alpha_R = \alpha_I = 0$, the equations can be readily solved to yield
\beqa
  s_x &=& - \frac{2 \eta_R}{\tilde{g}} \nonumber \\
  s_y &=& \frac{2 \eta_I}{\tilde{g}} \nonumber \\
  w &=& \pm \sqrt{1 - \frac{4 \eta^2}{\tilde{g}^2}}.
  \label{eq:sol a=0}
\eeqa
These solutions can be valid only for $\tilde{g} \geq 2 \eta/N \equiv g^*_1$.

In the case where $w=0$, we can use the spin conservation to parametrize the spin as $s_x = N \cos \theta, s_y = - N \sin \theta$. (\ref{eq:SS l0 D0 w}) then becomes
\beqa
  \frac{\alpha_I}{\alpha_R} = \tan \theta.
  \label{eq:tan}
\eeqa
Next we can express $\alpha_R,\alpha_I$ from (\ref{eq:SS l0 D0 aR},\ref{eq:SS l0 D0 aI}) as
\beqa
  \alpha_R &=& \frac{1}{2(\Delta_{\rm ph}^2 + \kappa^2)} \left[ \kappa (\tilde{g} N \sin \theta + 2 \eta_I) - \Delta_{\rm ph} (\tilde{g} N \cos \theta + 2 \eta_R) \right] \nonumber \\
  \alpha_I &=& -\frac{1}{2(\Delta_{\rm ph}^2 + \kappa^2)} \left[ \Delta_{\rm ph} (\tilde{g} N \sin \theta + 2 \eta_I) + \kappa (\tilde{g} N \cos \theta + 2 \eta_R) \right].
  \label{eq:alpha tan}
\eeqa
Substituting these expressions to (\ref{eq:tan}) yields the condition for the angle $\theta$ which determines the solution and can be found numerically. In order to proceed further analytically, we put $\eta_I=0$. The equation (\ref{eq:tan}) can then be cast in the form
\beq
  z + \kappa \cos \theta = \Delta_{\rm ph} \sin \theta
  \label{eq:gon eq}
\eeq
or equivalently
\beq
  (\kappa^2 + \Delta_{\rm ph}^2) \cos^2 \theta + 2 \kappa z \cos \theta + z^2 - \Delta_{\rm ph}^2 = 0
  \label{eq:cos eq}
\eeq
with solutions
\beq
  \cos \theta = \frac{-2 \kappa z \pm \sqrt{4 \kappa^2 z^2 - 4(z^2 - \Delta_{\rm ph}^2)(\kappa^2 + \Delta_{\rm ph}^2)} }{2(\kappa^2 + \Delta_{\rm ph}^2)},
  \label{eq:sol cos}
\eeq
where $z = \kappa \tilde{g} N / (2 \eta_R)$. It is easy to understand the structure of solutions from the graphical representation of (\ref{eq:gon eq}), which is shown in Fig. \ref{fig:gon}. The left (right) hand side of (\ref{eq:gon eq}) is represented by solid blue (red) line respectively. Assuming $\Delta_{\rm ph} > 0$, \fig{\ref{fig:gon}a} (\fig{\ref{fig:gon}b}) shows a situation for $\tilde{g} < g_1^*$ ($\tilde{g} > g_1^*$) respectively. If $\tilde{g} < g_1^*$, one of the solutions is negative, while both solutions are positive for $\tilde{g} > g_1^*$.

When tuning $\tilde{g}$, the non negativity of the discriminant in (\ref{eq:sol cos}) determines the maximal coupling $g^*_2$ up to which the solutions (\ref{eq:sol cos}) can be found. It reads
\beq
  g^*_2 = \frac{2 \eta_R}{N} \sqrt{1 + \left( \frac{\Delta_{\rm ph}}{\kappa} \right)^2}.
  \label{eq:g crit 2}
\eeq
Note, that the region $\tilde{g} \in [g^*_1, g^*_2]$ admits both solutions (\ref{eq:sol a=0}) and (\ref{eq:alpha tan},\ref{eq:sol cos}) indicating a bistable behaviour. In what follows we refer to the points $g_1^*,g_2^*$ as transition points.

\begin{center}
	\begin{figure}[t!p]
  	\includegraphics[width=13cm]{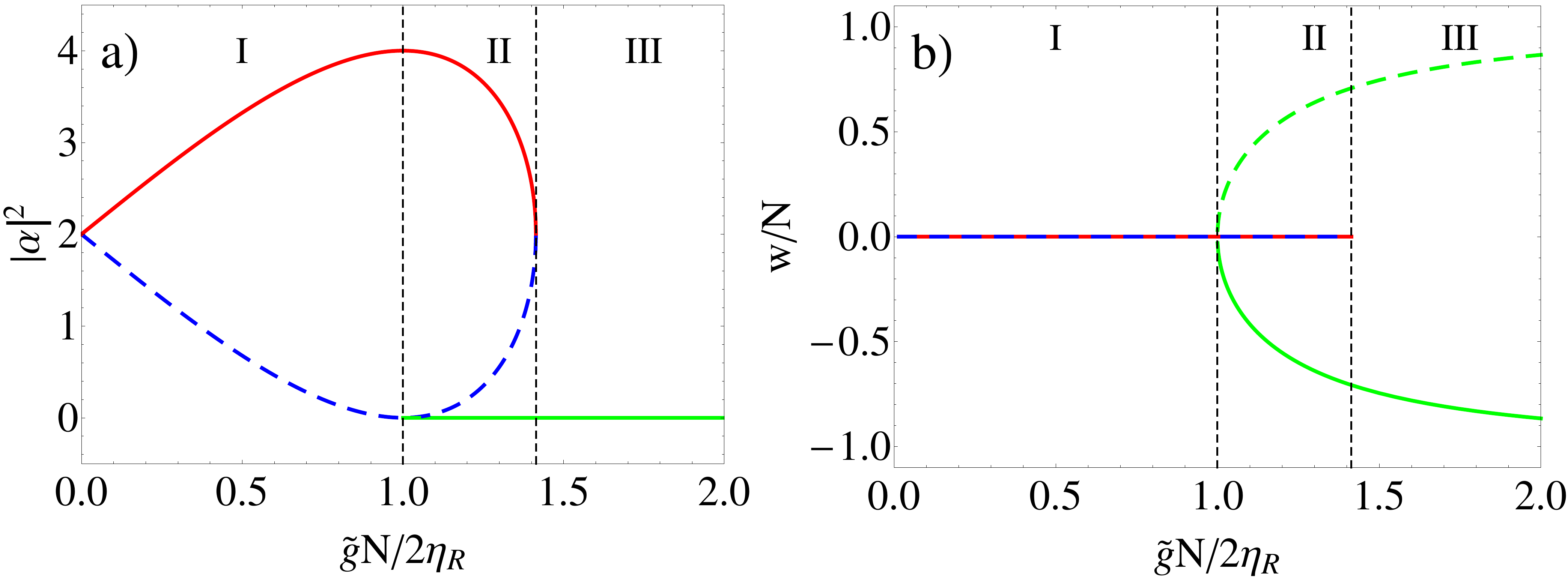} 
  	\caption{(Color online) Steady state phase diagram. I, II and III denote the regions of $\tilde{g}<g^*_1$, $g^*_1 \leq \tilde{g} \leq g^*_2$ and $\tilde{g} > g^*_2$ respectively. Stable (unstable) steady state solutions are indicated by solid (dashed) lines. (a) $|\alpha|^2$ and (b) $w$ as functions of the coupling $\tilde{g}$. Region I exhibits two possible solutions corresponding to the two solutions of (\ref{eq:alpha tan}). Region II exhibits the coexistence of the solutions (\ref{eq:alpha tan}) together with the solution (\ref{eq:sol a=0}) indicating bistability. The line colors are used as eye guide to help to identify corresponding $|\alpha|^2$ and $w$ solutions. Parameter values used are $\eta_I = 0, \Delta_{\rm ph}/\eta_R=0.5, \kappa/\eta_R=0.5$.}
  	\label{fig:stability}
 	\end{figure}
\end{center}

~\\
\emph{Stability study} \\
The stability analysis of the steady state solutions (\ref{eq:sol a=0}) and (\ref{eq:alpha tan},\ref{eq:sol cos}) is performed in a standard way by linearizing (\ref{eq:EoMs MF pump real}) around the solutions, i.e. expressing the variables as $v=\bar{v} + \delta v$, where $\bar{v}$ denotes the steady state solution. Formally this yields the linearized equations of motion
\beq
  \dot{\delta v} = M \delta v + b,
  \label{eq:stability}
\eeq
where $\delta v = (\delta \alpha_R, \delta \alpha_I, \delta s_x, \delta s_y, \delta w)^T$. Instability of the solutions is indicated by the positivity of the real part of the maximum eigenvalue of $M$, see Appendix \ref{app:General Stability} for details. It is noteworthy, that in any steady state solution we have either $\bar{w}=0$ or $\bar{\alpha}_R = \bar{\alpha}_I=0$ and consequently the characteristic polynomial of the matrix $M$ becomes
\beq
  |M-y \mathds{1}| = y p(y^4),
\eeq
where $p(y^4)$ is some polynomial which is 4th order in $y$. We thus always have one eigenvalue $y=0$, which is simply the consequence of the spin conservation law (\ref{eq:spin conservation}). Examining numerically the negativity of real part of the roots of $p(y^4)$ we identify stable and unstable solutions. These are depicted by solid and dashed lines respectively in steady state phase diagram Fig. \ref{fig:stability}, where we plot $|\alpha|^2$ and $w$ as functions of the coupling $\tilde{g}$. The region II exhibits bistable behaviour, whereas regions I and III admit only a single stable solution.

We have also verified, by numerically solving the dynamical equations (\ref{eq:EoMs MF pump real}), that they indeed evolve into the steady state solutions (\ref{eq:sol a=0},\ref{eq:alpha tan},\ref{eq:sol cos}) (not shown). In the next two subsections we investigate how the inclusion of the non-zero atomic detuning $\Delta_{\rm at}$ and non-zero coupling $\lambda$ terms modifies the $\Delta_{\rm at} = \lambda = 0$ solution.

\begin{center}
	\begin{figure}[t!p]
	\hspace*{0cm}
  	\includegraphics[width=13cm]{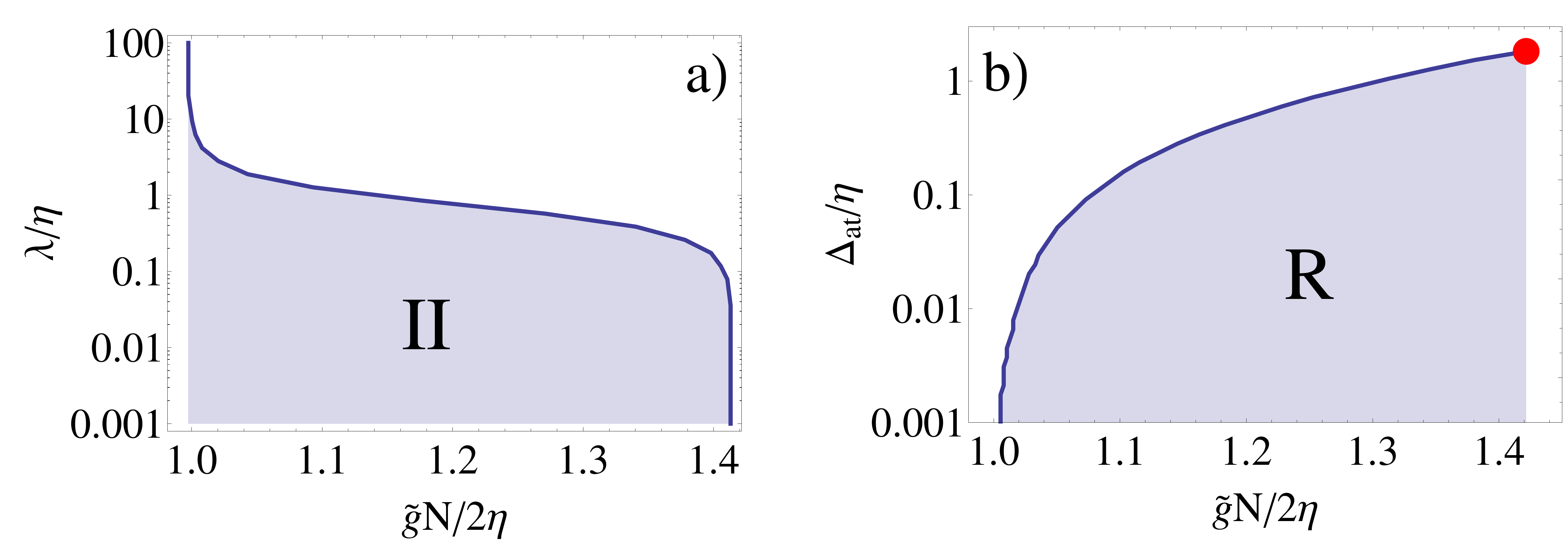} 
  	\caption{(Color online) (a) $\lambda \neq 0$ regime: the shaded area represents the shrinking of the $[g_1^*,g_2^*]$ region (denoted as region II, see also Fig. \ref{fig:stability}) with increasing $\lambda$. (b) $\lambda = 0$ regime: the shaded area corresponds to the region of $\tilde{g}$ exhibiting four solutions for a given $\tilde{g}$ (denoted as region R, see also Fig. \ref{fig:Delta_grid}a,b) as $\Delta_{\rm at}$ is increased. The critical value of $\Delta_{\rm at}$ at which the four simultaneous solutions cease to exist is represented by the red circle.}
  	\label{fig:g2 to g1}
 	\end{figure}
\end{center}
\subsubsection{$\lambda \neq 0$ regime}

In order to shed light on the effect of the $\lambda$ term independently of the $\Delta_{\rm at}$ term, we keep $\Delta_{\rm at}=0$. From (\ref{eq:EoMs MF pump real}) the steady state equations read
\begin{subequations}
\label{eq:SS lneq0 D0}
\begin{align}
   -\kappa \alpha_R + \left(\Delta_{\rm ph} + \lambda w \right) \alpha_I &= - \frac{\tilde{g}}{2} s_y + \eta_I \label{eq:SS lneq0 D0 aR} \\
    \left( \Delta_{\rm ph} + \lambda w \right) \alpha_R + \kappa \alpha_I &= - \frac{\tilde{g}}{2} s_x  - \eta_R \label{eq:SS lneq0 D0 aI} \\
  \lambda |\alpha|^2 s_y &= -\tilde{g} w \alpha_I \label{eq:SS lneq0 D0 sx} \\
    \lambda |\alpha|^2 s_x &= \tilde{g} w \alpha_R \label{eq:SS lneq0 D0 sy} \\ 
  \alpha_R s_y &=  - \alpha_I s_x. \label{eq:SS lneq0 D0 w}  
\end{align}
\end{subequations}
Finding the steady state solution encompasses solving a 6th order polynomial equation, which in general can only be done numerically. Nevertheless, some information about the steady state solution can be obtained analytically as we now describe.

Substituting $s_x, s_y$ from (\ref{eq:SS lneq0 D0 sx},\ref{eq:SS lneq0 D0 sy}) to the spin conservation (\ref{eq:spin conservation}), we get the following condition for $w$
\beq
	w^2 = \frac{\lambda^2 |\alpha|^2}{\lambda^2 |\alpha|^2 + \tilde{g}^2} N^2.
	\label{eq:cond w2}
\eeq
Since $\lambda^2 |\alpha|^2/(\lambda^2 |\alpha|^2 + \tilde{g}^2) \leq 1$, the relation (\ref{eq:cond w2}) indicates, that there is no further instability ($w^2 > N^2$) when changing $\lambda$, i.e. there are also two transition points $g_1^*,g_2^*$ for $\lambda \neq 0$.

Next, one can find solutions for asymptotic values of the parameter $\lambda$. Clearly, for $\lambda \rightarrow 0$, one should recover the solutions of Sec. \ref{sec:linear}. On the other hand, for $\lambda \rightarrow \infty$, one can look for solution by substituting a perturbative expansion for all the variables of the form $v = \sum_{n=0}^\infty \lambda^{-n} v^{(n)}$. We provide the details of this expansion in Appendix \ref{app:lambda exp.}. In order to simplify the analytic expressions we take the imaginary part of the pump to be zero, $\eta_I = 0$. In this case, to leading order in $\lambda$, the solutions read $\alpha_R^{(0)} = \alpha_I^{(0)} = s_y^{(0)} = 0$ and
\begin{subequations}
\label{eq:sx0}
\begin{align}
   s_x^{(0)} &= -\frac{2 \eta}{\tilde{g}} \;\; {\rm for} \;\; \tilde{g}\geq \frac{2 \eta}{N} \label{eq:sx0 1}\\
   s_x^{(0)} &= \frac{\eta}{\tilde{g}} \left(1 - \sqrt{1+\frac{2 \tilde{g}^2 N^2}{\eta^2}}\right) \;\; {\rm for} \;\; \tilde{g}\leq \frac{2 \eta}{N}. \label{eq:sx0 2}
\end{align}
\end{subequations}
Interestingly, since the spin conservation implies $\left| s_x^{(0)} \right| \leq N$, the two solutions (\ref{eq:sx0 1}), (\ref{eq:sx0 2}) yield the same transition point $g_1^* = 2 \eta/N$. This is in contrast to the $\lambda=0$ case, where two distinct transition points $g_1^*, g_2^*$ were identified. Here, the solution (\ref{eq:sx0 1}) is valid for $\tilde{g} \geq 2\eta/N$, whereas (\ref{eq:sx0 2}) is valid for $\tilde{g} \leq 2\eta/N$. This indicates that the transition point $g_2^*$ approaches $g_1^*$ as $\lambda$ is increased until they become identical in the $\lambda \rightarrow \infty$ limit. In other words, the region II of the phase diagram \fig{\ref{fig:stability}} is shrinking to zero width as $\lambda$ is increased. For the intermediate values of $\lambda$, the values of $g_1^*,g_2^*$ can be found numerically (selecting only the physically meaningful solutions of the 6th order polynomial, corresponding to the real values of $w$). We plot the shrinking of the $[g_1^*,g_2^*]$ region in \fig{\ref{fig:g2 to g1}a}.
\begin{center}
	\begin{figure}[t!p]
	\hspace*{0cm}
  	\includegraphics[width=13cm]{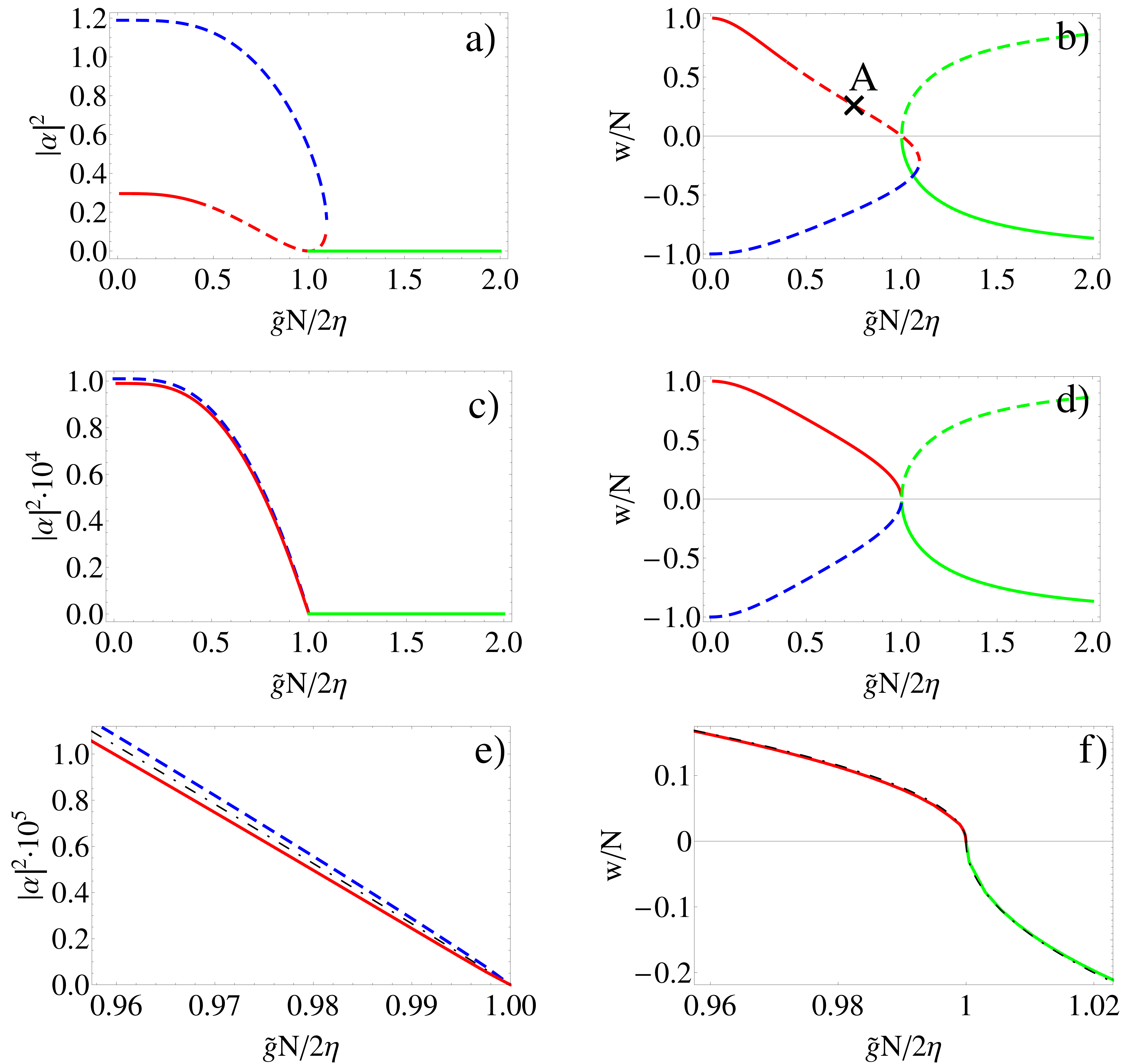} 
  	\caption{(Color online) The cavity field $|\alpha|^2$ and the expectation value $w$ of the spin as functions of the coupling strength $\tilde{g}$ for various values of the nonlinearity $\lambda$: $\lambda/\eta = 1.3$ in (a,b) and $\lambda/\eta=100$ in (c,d). Stable (unstable) solutions are shown as solid (dashed) lines. Panels (e,f) show a magnification of the data shown in (c,d) in the vicinity of the critical point $g_1^*$. The dashed dotted lines represent the analytic scaling solutions (\ref{eq:a2}) and (\ref{eq:w0}) respectively. In all plots the line colors are used as eye guide to help to identify the corresponding $|\alpha|^2$ and $w$ solutions.}
  	\label{fig:lambda_grid}
 	\end{figure}
\end{center}
\begin{center}
	\begin{figure}[t!p]
	\hspace*{0cm}
  	\includegraphics[width=13cm]{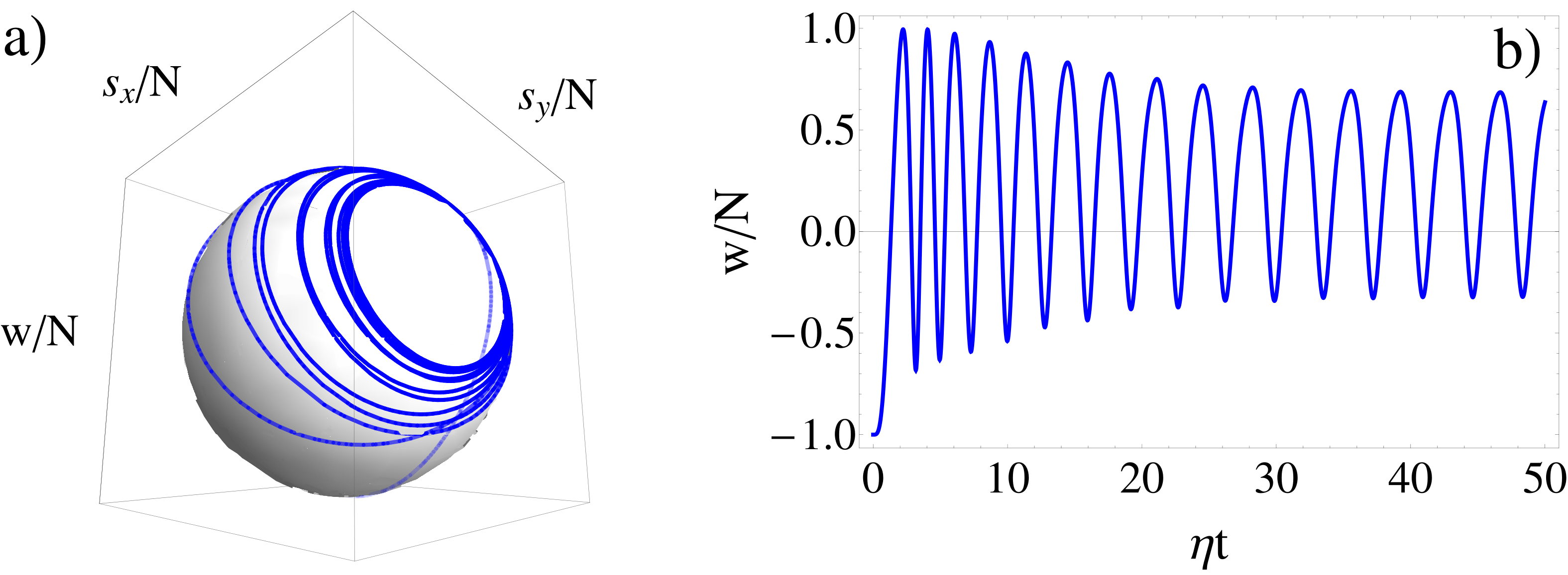} 
  	\caption{(Color online) (a) Limit cycle corresponding to $\lambda/\eta=1.3$ and $\tilde{g}N/2\eta = 0.75$ (point A in \fig{\ref{fig:lambda_grid}}b) represented by the evolution of the spin on the Bloch sphere. Here the direction of time corresponds to the increasing opacity of the spin evolution line. (b) Evolution of $w$ as a function of time. We take the initial (final) time to be the same in (a) and (b).}
  	\label{fig:limit_cycle}
 	\end{figure}
\end{center}
In \fig{\ref{fig:lambda_grid}} we show the phase diagrams for $|\alpha|^2$ and $w$ for increasing values of $\lambda$. It can be seen from \fig{\ref{fig:lambda_grid}a,b}, that for some values of $\lambda$, there is a region with no apparent stable solution. In non-linear systems, this is typically a signature of the appearance of limit cycles which occur through a Hopf bifurcation as the system leaves the stable fixed point by changing the coupling $\tilde{g}$ \cite{Strogatz_1994}. We represent a limit cycle corresponding to the point A in \fig{\ref{fig:lambda_grid}b} as the time evolution of the global spin in \fig{\ref{fig:limit_cycle}}.

One can carry the analysis further in the large $\lambda$ limit and look for asymptotic solutions in the vicinity of the unique transition point $g_1^*$. The first non-trivial contribution to $\alpha$ is of order $\lambda^{-1}$ and given by (see Appendix \ref{app:lambda exp.})
\beq
	\left| \alpha \right|^2 = \frac{1}{\lambda^2} \frac{1}{27}\left( g_1^* - \tilde{g} \right)\left( \frac{10\eta}{N} + 13\tilde{g} \right) + O((g_1^* -\tilde{g})^{\frac{3}{2}}),
	\label{eq:a2}
\eeq
which is valid for $\tilde{g} \leq g_1^*$ as the solution for $\tilde{g} > g_1^*$ is trivial ($\alpha=0$). The scaling of the spin observables, say $w$, is simply obtained from the expansion of (\ref{eq:sx0}), which read
\begin{subequations}
\label{eq:w0}
\begin{align}
   w^{(0)} &= \pm  N \sqrt{\frac{N}{\eta}} \sqrt{\tilde{g}-g_1^*}+ O((\tilde{g}-g_1^*)^{\frac{3}{2}}) \;\; {\rm for} \;\; \tilde{g}\geq g_1^* \label{eq:w0 1}\\
   w^{(0)} &= \pm N \sqrt{\frac{N}{3 \eta}} \sqrt{g_1^*-\tilde{g}}  + O((g_1^* -\tilde{g})^{\frac{3}{2}})  \;\; {\rm for} \;\; \tilde{g}\leq g_1^*, \label{eq:w0 2}
\end{align}
\end{subequations}
where the sign $\pm$ depends on the branch of $w$ considered. Note that the scaling $(\tilde{g}-g_1^*)^{1/2}$ is characteristic of the Ising type model with $\mathbb{Z}_2$ symmetry and $g_1^*$ is the phase transition critical point. We compare the asymptotic solutions (\ref{eq:a2}),(\ref{eq:w0}) with the solutions obtained by numerically solving (\ref{eq:SS lneq0 D0}) in \fig{\ref{fig:lambda_grid}e,f}.

\begin{center}
	\begin{figure}[t!p]
	\hspace*{0cm}
  	\includegraphics[width=13cm]{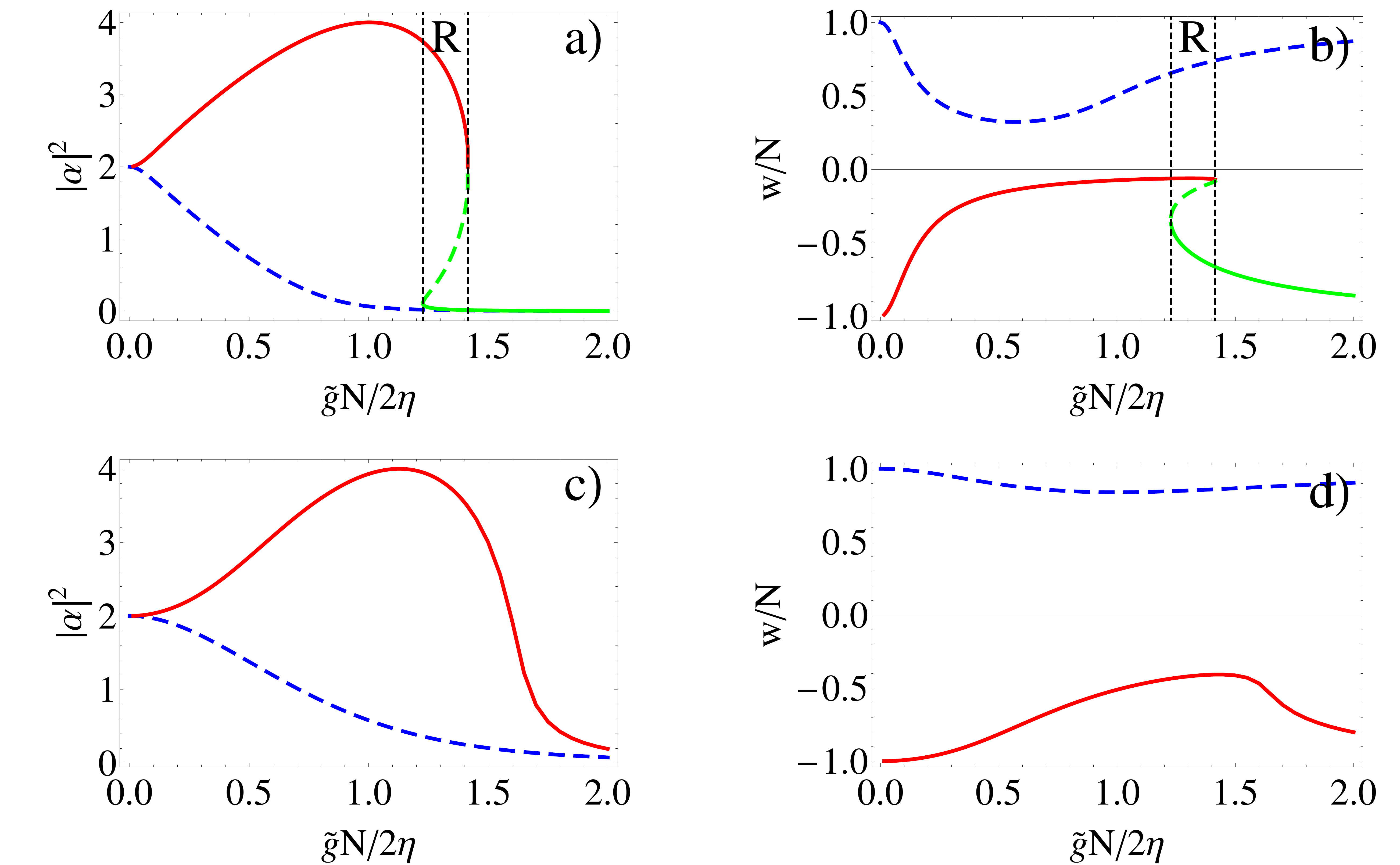} 
  	\caption{(Color online) $|\alpha|^2$ and $w$ for $\Delta_{\rm at}/\eta = 0.5$ (a,b) and $\Delta_{\rm at}/\eta = 3.9$ (c,d). R in (a,b) denotes a region exhibiting four distinct solutions. Stable (unstable) solutions are shown as solid (dashed) lines. The line colors are used as eye guide to help to identify the corresponding $|\alpha|^2$ and $w$ solutions. Parameter values used are $\eta_I = 0, \Delta_{\rm ph}/\eta_R=0.5, \kappa/\eta_R=0.5$.}
  	\label{fig:Delta_grid}
 	\end{figure}
\end{center}
\subsubsection{$\lambda=0$ regime with $\Delta_{\rm at} \neq 0$}

Our next aim is to explore the effect of the $\Delta_{\rm at}$ term on the solution in the absence of the non-linear $\lambda$ term. In this case, the structure of the solutions is dictated by the 4th order $w$ polynomial (\ref{eq:poly w}) and it is straightforward to obtain the solutions numerically. For small $\Delta_{\rm at}$, there is a range of $\tilde{g}$-values, which admits four solutions (corresponding to four real solutions of (\ref{eq:poly w})), Fig. \ref{fig:Delta_grid}a,b. As $\Delta_{\rm at}$ is increased, this region eventually disappears leaving us with only two solutions for all values of $\tilde{g}$. The latter limit can be simply understood from the MF equations (\ref{eq:EoMs MF pump real SS}). In the large $\Delta_{\rm at}$ limit, the leading contribution comes from (\ref{eq:EoMs MF pump real SS sx},\ref{eq:EoMs MF pump real SS sy}) with the trivial solutions $s_x = s_y = 0$ implying $w/N = \pm 1$. The evolution of the solutions towards this large $\Delta_{\rm at}$ limit can be seen rather clearly from Fig. \ref{fig:Delta_grid}d. Next we have determined numerically the size of the region of $\tilde{g}$ admitting four solutions as a function of $\Delta_{\rm at}$. This is shown in Fig. \ref{fig:g2 to g1}b. Clearly, there is some maximal $\Delta_{\rm at}$ after which there are only two possible solutions, as discussed. This limiting value is represented by the red circle in Fig. \ref{fig:g2 to g1}b. Note, that the inclusion of the $\Delta_{\rm at}$ term also lifts the transition point $g_1^*$, i.e. that all solutions are smooth in the vicinity of $g_1^*$.

\subsection{Multiple cavities}
\label{sec:2D}

In this section we seek the generalization of the single cavity case to higher dimensional geometries. For concreteness, we consider a 2D square  geometry depicted schematically in \fig{\ref{fig:Scheme}a}. 

Before diving in the details of the analysis, we motivate this section by asking whether a 2D square geometry offers MF solutions which are qualitatively different from the 1D case studied above. For example, it was shown in \cite{Lee_2011} in the context of laser driven and interacting Rydberg gases on a square lattice, that a homogeneous system admits a MF solution that breaks the lattice symmetry and exhibits antiferromagnetic (AF) order.

We start our discussion by deriving the 2D MF equations, which follow from the operator equations of motion (\ref{eq:op EoMs}) given by the Hamiltonian (\ref{eq:H eff RWA 2D}). Repeating the argument yielding (\ref{eq:restr}) leads to the same conclusion, namely that in the absence of the cavity pump the system posses only a trivial solution, where the cavity modes are empty and all the spins are down (up). We first focus on the situation without the non-linear term, $\lambda = 0$. In the following, we consider the cavity pump $\eta$, the cavity decay $\kappa$ and the spin decay $\gamma$ to be the same for all cavity modes and all spins respectively (the motivation for adding the spin decay will become clear shortly). We obtain the set of MF equations
\begin{subequations}
\label{eq:MF 2D}
\begin{align}
	i \dot{\alpha}_{\bar{i}} &= \left( \Delta_{\rm ph}^{\bar{i}} - i \kappa + \lambda w_{\bar{i} \nu} \right) \alpha_{\bar{i}} + \tilde{g}_a s_{\bar{i} \nu} + \eta + \lambda \beta_{\nu} \left( w_{\bar{i} \nu} - \mathds{1} \right) \label{eq:MF 2D a} \\
	i \dot{\beta}_{\bar{\nu}} &= \left( \Delta_{\rm ph}^{\bar{\nu}} - i \kappa + \lambda w_{i \bar{\nu}} \right) \beta_{\bar{\nu}} + \tilde{g}_b s_{i \bar{\nu}} + \eta + \lambda \alpha_{i} \left( w_{i \bar{\nu}} - \mathds{1} \right) \label{eq:MF 2D b}\\
	i \dot{s}_{\bar{i} \bar{\nu}} &= \left( \Delta_{\rm at} -i\frac{\gamma}{2}  + 2 \lambda \left( \alpha^*_{\bar{i}} + \beta^*_{\bar{\nu}} \right) \left( \alpha_{\bar{i}} + \beta_{\bar{\nu}} \right) \right) s_{\bar{i} \bar{\nu}} - \left( \tilde{g}_a \alpha_{\bar{i}} + \tilde{g}_b \beta_{\bar{\nu}} \right) w_{\bar{i} \bar{\nu}} \label{eq:MF 2D s} \\
	i \dot{w}_{\bar{i} \bar{\nu}} &= 2 \left[ s^*_{\bar{i} \bar{\nu}} \left( \tilde{g}_a \alpha_{\bar{i}} + \tilde{g}_b \beta_{\bar{\nu}} \right) - {\rm h.c.} \right] - i \gamma \left( w_{\bar{i} \bar{\nu}} + \mathds{1} \right). \label{eq:MF 2D sz}
\end{align}
\end{subequations}

Here, we allowed for the couplings $\tilde{g}$ and the photon detunings $\Delta_{\rm ph}$ to be different for rows and columns while taking the remaining parameters to be the same for all cavities.

Clearly, the 2D system provides higher tunability by enlarging the parameter space. In the following, we study the problem considering two different perspectives, namely a homogeneous and cluster-MF ansatz.

\subsubsection{Square array with homogeneous MF}
\label{sec:asym array}

In order to further simplify the MF equations (\ref{eq:MF 2D}), it is reasonable to use the typical MF ansatz, namely that, due to translational symmetry in either direction of the 2D array (along rows or columns), the corresponding cavity fields are the same ($\alpha_{\bar{i}} = \alpha$, $\beta_{\bar{\nu}} = \beta$ for all $\bar{i},\bar{\nu}$) as well as the spins, $s_{\bar{i}\bar{\nu}} = s,w_{\bar{i}\bar{\nu}} = w$. We can now proceed along similar lines as in Sec. \ref{sec:linear} in order to characterize the solutions. In analogy to Sec. \ref{sec:linear} we start with $\Delta_{\rm at} = \lambda = 0$ and we neglect the spin decay for the moment, $\gamma=0$. Considering $\Delta_{\rm ph}^{\bar{i}} \equiv \Delta_{\rm ph}^a$ and $\Delta_{\rm ph}^{\bar{\nu}} \equiv \Delta_{\rm ph}^b$ for all $\bar{i},\bar{\nu}$ to be the same along rows or columns, the steady state MF equations (\ref{eq:MF 2D}) simplify to
\begin{subequations}
\label{eq:MF 2D ass}
\begin{align}
	0 &= \left( \Delta_{\rm ph}^{a} - i \kappa  \right) \alpha + \tilde{g}_a N_C s + \eta \label{eq:MF 2D ass a} \\
	0 &= \left( \Delta_{\rm ph}^{b} - i \kappa \right) \beta + \tilde{g}_b N_R s +\eta \label{eq:MF 2D ass b} \\
	0 &= \left( \tilde{g}_a \alpha + \tilde{g}_b \beta \right) w \label{eq:MF 2D ass s} \\
	0 &= s^* \left( \tilde{g}_a \alpha + \tilde{g}_b \beta \right) - {\rm h.c.} \label{eq:MF 2D ass sz}.
\end{align}
\end{subequations}
Here $N_R$ ($N_C$) is the number of rows (columns) respectively. Since the fields $\beta$ and $\alpha$ share the same spin $s$, $\beta$ can be given directly in terms of the field $\alpha$. The situation is then essentially equivalent to the single cavity case analyzed in Sec. \ref{sec:Single cavity}, though with some extra tunability provided by larger number of parameters. For example, one can find the transition point $g_1^*$ by combining (\ref{eq:MF 2D ass a}) and (\ref{eq:MF 2D ass b}) with $\tilde{g}_a \alpha + \tilde{g}_b \beta = 0$, which is one possible solution of (\ref{eq:MF 2D ass s}). The transition point then corresponds to the situation where $w=0$, i.e.
\beq
  |s|^2 = \frac{1}{4} = |\eta|^2 \frac{\left( \Delta_{\rm ph}^b \tilde{g}_a + \Delta_{\rm ph}^a \tilde{g}_b \right)^2 + \left(\tilde{g}_a + \tilde{g}_b \right)^2 \kappa^2 }{\left( \Delta_{\rm ph}^b \tilde{g}_a^2 N_C + \Delta_{\rm ph}^a \tilde{g}_b^2 N_R \right)^2 + \left(\tilde{g}_a^2 N_C + \tilde{g}_b^2 N_R \right)^2 \kappa^2 }.  
\eeq
One can then identify the critical value of e.g. $\tilde{g}_a$ for all other parameters fixed. As a consistency check, it is easy to verify that one recovers the single cavity expression $g_1^* = 2|\eta|/N$ by omitting all the "$b$" variables and setting $N_R=1$ and $N_C = N$, the number of spins.

\subsubsection{Square array with cluster-MF}
\label{sec:symm array}

In the previous section, we have used the homogeneous MF ansatz and found that the solutions correspond effectively to a single cavity case with no intriguing spin configurations. However, it is known and was shown e.g. in \cite{Lee_2011}, that a suitable MF ansatz can lead to a non-trivial configuration (such as antiferromagnetic ordering) even if the steady state MF equations are completely symmetric under exchange of the spin variables. Here we will consider the simplest possible case of such ansatz where a cluster is formed by two adjacent inequivalent spins $s_1,w_1$ and $s_2,w_2$. On the other hand, since the fields $\alpha$ and $\beta$ couple to both $s_1$ and $s_2$ in the same way, we take $\alpha,\beta$ to be the same along rows (columns) due to translational symmetry. In order to simplify the equations, we now take all the parameters to be the same along all rows and columns (i.e. we set $\tilde{g}_a = \tilde{g}_b = \tilde{g}$ and $\Delta_{\rm ph}^{\bar i} = \Delta_{\rm ph}^{\bar \nu} = \Delta_{\rm ph}$ for all $\bar{i}, \bar{\nu}$). Starting with the simplest case $\Delta_{\rm at} = \lambda = 0$, the MF equations (\ref{eq:MF 2D}) become
\begin{subequations}
\label{eq:MF 2D symm}
\begin{align}	
	0 &= \left( \Delta_{\rm ph} - i \kappa  \right) \alpha + \frac{\tilde{g} \sqrt{N}}{2} \left(s_1 + s_2 \right) + \eta \label{eq:MF 2D symm a} \\
	0 &= \left( \Delta_{\rm ph} - i \kappa \right) \beta + \frac{\tilde{g} \sqrt{N}}{2} \left(s_1 + s_2 \right) +\eta \label{eq:MF 2D symm b} \\
	0 &= i \frac{\gamma}{2} s_{\bar{j}} + \tilde{g} \left(\alpha + \beta \right) w_{\bar{j}} \label{eq:MF 2D symm s} \\
	0 &= 2 \tilde{g} \left[ s^*_{\bar{j}} \left( \alpha + \beta \right) - {\rm h.c.} \right] - i \gamma \left( w_{\bar{j}} + \mathds{1} \right) \label{eq:MF 2D symm sz},
\end{align}
\end{subequations}
where $\bar{j}=1,2$ labels the different spins of the cluster and we assumed that each field $\alpha,\beta$ couples to the same number of spins 1 and 2, hence the factor $1/2$ in the second term of (\ref{eq:MF 2D symm a}),(\ref{eq:MF 2D symm b}). Here $N$ denotes the total number of spins, i.e. there are $\sqrt{N}$ spins along rows and along columns. 

Until now we did not comment on the spin conservation in the 2D case. Going back to the most general situation, where every individual spin and cavity mode is described in terms of the corresponding MF variables $s_{\bar{i} \bar{\nu}},w_{\bar{i} \bar{\nu}}$ and $\alpha_{\bar{i}},\beta_{\bar{\nu}}$ respectively, the system evolves according to the full set of the 2D MF equations (\ref{eq:MF 2D}). One can verify by means of (\ref{eq:diff spin conservation}) that (\ref{eq:MF 2D}) actually imply both local
\beq
  w_{\bar{i} \bar{\nu}}^2 + 4 \left| s_{\bar{i} \bar{\nu}}^2 \right|^2 = 1, \;\; \forall \bar{i}, \bar{\nu}
  \label{eq:spin conservation local}
\eeq
and global spin conservation
\beq
  W^2 + 4 \left| \Sigma \right|^2 = N^2,
  \label{eq:spin conservation global}
\eeq
provided $\gamma=0$. In (\ref{eq:spin conservation global}) $W = w_{i\nu}$ and $\Sigma = s_{i \nu}$ are the global spin components. One should appreciate that (\ref{eq:spin conservation global}) actually prevents any spin configuration incompatible with it, including the AF order (which corresponds to $w_1 = - w_2$ when using the here considered cluster-MF). In order to see e.g. the AF order, such as in \cite{Lee_2011}, one needs to break the global spin conservation. This is achieved by the inclusion of the spin decay $\gamma$, which breaks both the local and global spin conservations (\ref{eq:spin conservation local}),(\ref{eq:spin conservation global}). 

Allowing for the spin decay, we can now ask, whether the solution of the cluster-MF equations (\ref{eq:MF 2D symm}) features any non-trivial spin configuration ($w_1 \neq w_2$). First, it follows from (\ref{eq:MF 2D symm a}) and (\ref{eq:MF 2D symm b}) that $\alpha=\beta$. Next, expressing for $s_{\bar{j}}$ from (\ref{eq:MF 2D symm s}) and substituting to (\ref{eq:MF 2D symm sz}), it can be shown that
\beq
  \frac{w_1+1}{w_1} = \frac{w_2+1}{w_2} \; \Rightarrow \; w_1 = w_2
\eeq
and $s_1 = s_2$ from (\ref{eq:MF 2D symm s}). Once again we find that the structure of the equations (\ref{eq:MF 2D symm}) reduces the problem effectively to the single cavity situation described by a simple set of variables $\alpha, s, w$ and this even when allowing for inequivalent spin configurations and the spin decay.

One might argue, that the equivalence to the single cavity case conjectured in sections \ref{sec:asym array} and \ref{sec:symm array} is an artefact of taking $\Delta_{\rm at}=\lambda=0$. Indeed, when comparing (\ref{eq:EoMs MF}) and (\ref{eq:MF 2D}), one can note that there is a \emph{qualitative} difference between the 1D and 2D situation. Specifically, there is an extra coupling between the $a$ and $b$ modes, the last term in (\ref{eq:MF 2D a}) and (\ref{eq:MF 2D b}). We have numerically verified that, in a general situation with variables $\alpha,\beta, s_{\bar{j}}, w_{\bar{j}}$, $\bar{j}=1,2$ and $\tilde{g}_a \neq \tilde{g}_b$, $\Delta_{\rm ph}^a \neq \Delta_{\rm ph}^b$, $\Delta_{\rm at} \neq 0 \neq \lambda$, the solutions always yield $s_1=s_2,w_1=w_2$. For completeness we have included in our numerical analysis also the special cases, where any possible combination of the following conditions can occur: $\tilde{g}_a = \tilde{g}_b$, $\Delta_{\rm ph}^a = \Delta_{\rm ph}^b$, $\Delta_{\rm at} = 0$, $\lambda=0$.

In summary, the MF equations (\ref{eq:MF 2D}) with the ansatz considered in sections \ref{sec:asym array} and \ref{sec:symm array} on a square lattice reduce to effectively one-dimensional description with no intriguing spin configurations. It would be desirable to perform a beyond MF study of the non-linear two-dimensional model in order to asses the true nature of the steady state and the corresponding spin and field configurations, including their mutual correlations. Also we did not fully exploit the possibilities offered by the proposed implementation, such as taking different geometries of the array or allowing for disordered coupling strengths, which we leave for further investigations.

\section{Conclusion}
\label{sec:Conclusion}

Motivated by the progress in integrated optical circuits, we have proposed a possible realization of a two-dimensional cavity array with trapped atoms, which is a promising scalable quantum architecture. We derived an effective description of the system in terms of Jaynes-Cummings like Hamiltonian with highly tunable parameters and extra non-linear terms. We then analyzed the dynamics of the system using a MF approach. We have found a rich behaviour including bistable regions, Ising like phase transition or occurrence of limit cycles through Hopf bifurcations. In the present setup, we have not found conceptual differences between the one and two-dimensional cases at the level of the MF description and with the geometry considered. 
We hope that the present work lays down grounds for future studies of the cavity arrays realized with integrated optical circuits. The problems which might be addressed in the future are e.g. going beyond MF description, accounting for more exotic geometries or studying effective spin physics as a low energy limit of the presented cavity Hamiltonian.

\section{Acknowledgments}
\label{sec:Acknowledgments}

We thank T. Pohl, J. Goette, D. Jukic and all members of the QuILMI consortium for fruitful discussions. J.M. would like to thank Matteo Marcuzzi for useful discussions. The research leading to these results has received funding from the European Research Council under the European Union's Seventh Framework Programme (FP/2007-2013) / ERC Grant Agreement No. 335266 (ESCQUMA) and the EU-FET grants HAIRS 612862 and QuILMI 295293.



\begin{thebibliography}{10}

\bibitem{Everitt_2005}
{\em Experimental Aspects of Quantum Computing}, edited by H.~O. Everitt
  (Springer, New York, 2005).

\bibitem{Bradley_2003}
R. Bradley {\it et~al.}, Rev. Mod. Phys. {\bf 75},  777  (2003).

\bibitem{Maschler_2008}
C. Maschler, I. Mekhov, and H. Ritsch, Eur. Phys. J. D {\bf 46},  545  (2008).

\bibitem{Dimer_2007}
F. Dimer, B. Estienne, A.~S. Parkins, and H.~J. Carmichael, Phys. Rev. A {\bf
  75},  013804  (2007).

\bibitem{Gammelmark_2011}
S. Gammelmark and K. M\o{}lmer, New J. Phys. {\bf 12},  053035  (2011).

\bibitem{Bakemeier_2012}
L. Bakemeier, A. Alvermann, and H. Fehske, Phys. Rev. A {\bf 85},  043821
  (2012).

\bibitem{Castanos_2012}
O. Casta\~nos, E. Nahmad-Achar, R. L\'opez-Pe\~na, and J.~G. Hirsch, Phys. Rev.
  A {\bf 86},  023814  (2012).

\bibitem{Bhaseen_2012}
M.~J. Bhaseen, J. Mayoh, B.~D. Simons, and J. Keeling, Phys. Rev. A {\bf 85},
  013817  (2012).

\bibitem{Strack_2011}
P. Strack and S. Sachdev, Phys. Rev. Lett. {\bf 107},  277202  (2011).

\bibitem{Buchhold_2013}
M. Buchhold, P. Strack, S. Sachdev, and S. Diehl, Phys. Rev. A {\bf 87},
  063622  (2013).

\bibitem{Domokos_2002}
P. Domokos and H. Ritsch, Phys. Rev. Lett. {\bf 89},  253003  (2002).

\bibitem{Zippilli_2004}
S. Zippilli, G. Morigi, and H. Ritsch, Phys. Rev. Lett. {\bf 93},  123002
  (2004).

\bibitem{Asboth_2005}
J.~K. Asb\'oth, P. Domokos, H. Ritsch, and A. Vukics, Phys. Rev. A {\bf 72},
  053417  (2005).

\bibitem{Black_2003}
A.~T. Black, H.~W. Chan, and V. Vuleti\ifmmode~\acute{c}\else \'{c}\fi{}, Phys.
  Rev. Lett. {\bf 91},  203001  (2003).

\bibitem{Baumann_2010}
K. Baumann, C. Guerlin, F. Brennecke, and T. Esslinger, Nature {\bf 464},  1301
   (2010).

\bibitem{Hartmann_2008a}
M.~J. Hartmann, F.~G. S.~L. Brandao, and M.~B. Plenio, Laser \& Photon. Rev.
  {\bf 2},  527  (2008).

\bibitem{Vuckovic_2001}
J. Vu\ifmmode \check{c}\else \v{c}\fi{}kovi\ifmmode~\acute{c}\else \'{c}\fi{},
  M. Lon\ifmmode~\check{c}\else \v{c}\fi{}ar, H. Mabuchi, and A. Scherer, Phys.
  Rev. E {\bf 65},  016608  (2001).

\bibitem{Vuckovic_2003}
J. Vu\v{c}kovi\'{c} and Y. Yamamoto, Appl. Phys. Lett. {\bf 82},  2347  (2003).

\bibitem{Marshall_2009}
G.~D. Marshall {\it et~al.}, Opt. Express {\bf 17},  12546  (2009).

\bibitem{Crespi_2011}
A. Crespi {\it et~al.}, Nat. Comm. {\bf 2},  566  (2011).

\bibitem{Bromberg_2009}
Y. Bromberg, Y. Lahini, R. Morandotti, and Y. Silberberg, Phys. Rev. Lett. {\bf
  102},  253904  (2009).

\bibitem{Keil_2011}
R. Keil {\it et~al.}, Phys. Rev. Lett. {\bf 107},  103601  (2011).

\bibitem{Grafe_2014}
M. Gr\"{a}fe {\it et~al.}, Nat. Phot. {\bf 8},  791  (2014).

\bibitem{Peruzzo_2010}
A. Peruzzo {\it et~al.}, Science {\bf 329},  1500  (2010).

\bibitem{Weimann_2015}
S. Weimann {\it et~al.}, arXiv:1508.00033  (2015).

\bibitem{Lebugle_2015}
M. Lebugle {\it et~al.}, arXiv:1501.01764  (2015).

\bibitem{Marshall_2006}
G. Marshall, M. Ams, and M. Withford, Opt. Lett. {\bf 31},  2690  (2006).

\bibitem{Thiel_2015}
M. Thiel, G. Flachenecker, and W. Schade, Opt. Lett. {\bf 40},  1266  (2015).

\bibitem{Derntl_2014}
C. Derntl {\it et~al.}, Opt. Express {\bf 22},  22111  (2014).

\bibitem{Potts_2016}
C.~A. Potts {\it et~al.}, arXiv:1601.03344  (2016).

\bibitem{Jaynes_1963}
E. Jaynes and F. Cummings, Proc. IEEE {\bf 51},  89  (1963).

\bibitem{Vogel_1989}
W. Vogel and D.-G. Welsch, Phys. Rev. A {\bf 40},  7113  (1989).

\bibitem{SebaweAbdalla_1990}
M. Sebawe~Abdalla, M. Ahmed, and A.-S. Obada, Physica A {\bf 162},  215
  (1990).

\bibitem{Vogel_1995}
W. Vogel and R.~L. d.~M. Filho, Phys. Rev. A {\bf 52},  4214  (1995).

\bibitem{Joshi_2000}
A. Joshi, Phys. Rev. A {\bf 62},  043812  (2000).

\bibitem{Sivakumar_2000}
S. Sivakumar, J. Phys. A {\bf 33},  2289  (2000).

\bibitem{Budini_2003}
A.~A. Budini, R.~L. de~Matos~Filho, and N. Zagury, Phys. Rev. A {\bf 67},
  033815  (2003).

\bibitem{Sivakumar_2004}
S. Sivakumar, International Journal of Theoretical Physics {\bf 43},  2405
  (2004).

\bibitem{Singh_2011}
S. Singh and Amrita, International Journal of Theoretical Physics {\bf 51},
  838  (2011).

\bibitem{Walls_1994}
{\em Quantum Optics}, edited by D. Walls and G.~J. Milburn (Springer, Berlin,
  1994).

\bibitem{Koch_2009}
J. Koch and K. Le~Hur, Phys. Rev. A {\bf 80},  023811  (2009).

\bibitem{Emary_2003}
C. Emary and T. Brandes, Phys. Rev. E {\bf 67},  066203  (2003).

\bibitem{Strogatz_1994}
{\em Nonlinear Dynamics and Chaos}, edited by S.~H. Strogatz (Perseus Books,
  Reading, Massachussets, 1994).

\bibitem{Lee_2011}
T.~E. Lee, H. H\"affner, and M.~C. Cross, Phys. Rev. A {\bf 84},  031402
  (2011).

\end{thebibliography}


\newpage

\begin{appendices}


\section{Adiabatic elimination}
\label{app:Ad.el}

Here we derive the effective Hamiltonian resulting from the adiabatic elimination of the state $\ket{e}$ and starting with the full system Hamiltonian given in (\ref{eq:H RWA 2D}), which we rewrite in the hard-core boson representation with annihilation operators $b_{\mu,\bar{i} \bar{\nu}}$, $\mu = e,s,g$ for the atom at position $\bar{i} \bar{\nu}$
\beqa
	H &=& \omega_i a_i^\dag a_i + \omega_\nu b_\nu^\dag b_\nu + \omega_e b^\dag_{e,i\nu} b_{e,i\nu} + \omega_s b^\dag_{s,i\nu} b_{s,i\nu} \nonumber \\
	&& + \Omega(b^\dag_{e,i\nu} b_{s,i\nu} {\rm e}^{-i \omega_T t} + b^\dag_{s,i\nu} b_{e,i\nu} {\rm e}^{i \omega_T t}) \nonumber \\
	&& + g (b^\dag_{e,i\nu} b_{g,i\nu} (a_i + b_\nu) + {\rm h.c.}).
	\label{eq:H RWA 2D hard core}
\eeqa

First, we determine the transformation to the rotating frame. We consider a general unitary transformation
\beq
  U = {\rm e}^{-it \left( \alpha_i a_i^\dag a_i + \beta_\nu b_\nu^\dag b_\nu + \gamma_{i\nu} b_{e,i\nu}^\dag b_{e,i\nu} + \epsilon_{i\nu} b_{s,i\nu}^\dag b_{s,i\nu} \right)},
\eeq
where $\alpha,\beta,\gamma,\epsilon$ are some arbitrary frequencies. The requirement of eliminating all explicit time dependencies in \eq{\ref{eq:H RWA 2D hard core}} leads to the conditions
\beqa
  \gamma_{\bar{i} \bar{\nu}} &=& \omega_T + \epsilon_{\bar{i} \bar{\nu}} \nonumber \\
  \alpha_{\bar{i}} &=& \gamma_{\bar{i} \bar{\nu}} \nonumber \\
  \beta_{\bar{\nu}} &=& \gamma_{\bar{i} \bar{\nu}}.
  \label{eq:cond RWA 2D}
\eeqa
Note that in general, it is not possible to bring the levels $\ket{g}$ and $\ket{s}$ to degeneracy for all atoms since $\omega_s+\omega_T = \omega_{\bar{i}} = \omega_{\bar{\nu}}$ cannot be satisfied for all $\bar{i}$ and $\bar{\nu}$ at the same time. On the other hand, one has a freedom in the choice of frequencies $\alpha-\epsilon$, provided the conditions (\ref{eq:cond RWA 2D}) are satisfied. We adopt the following choice
\beqa
  \alpha_{\bar{i}} &=& \beta_{\bar{\nu}} = \gamma_{\bar{i} \bar{\nu}} = \omega_{\rm aux} \nonumber \\
  \epsilon_{\bar{i} \bar{\nu}} &=& \omega_{\rm aux} - \omega_T,
\eeqa
where $\omega_{\rm aux}$ is an arbitrary auxiliary frequency. The Hamiltonian (\ref{eq:H RWA 2D}) then becomes
\beqa
	H &=& \Delta_i a_i^\dag a_i + \Delta_\nu b_\nu^\dag b_\nu + \Delta_e b^\dag_{e,i\nu} b_{e,i\nu} + \Delta_s b^\dag_{s,i\nu} b_{s,i\nu} \nonumber \\
	&& + \Omega(b^\dag_{e,i\nu} b_{s,i\nu} + b^\dag_{s,i\nu} b_{e,i\nu}) + g (b^\dag_{e,i\nu} b_{g,i\nu} (a_i + b_\nu) + {\rm h.c.}),
	\label{eq:H RWA 2D rot}
\eeqa
where $\Delta_x = \omega_x - \omega_{\rm aux}$, $x=\bar{i},\bar{\nu},e$ and $\Delta_s = \omega_s - (\omega_{\rm aux} - \omega_T)$. This leads to the equations of motion
\beqa
	i \dot{b}_{e,\bar{i}\bar{\nu}} &=& \Delta_e b_{e,\bar{i}\bar{\nu}} + \Omega b_{s,\bar{i}\bar{\nu}} + g b_{g,\bar{i}\bar{\nu}}(a_{\bar{i}} + b_{\bar{\nu}}) \nonumber \\
	i \dot{b}_{g,\bar{i}\bar{\nu}} &=& g b_{e,\bar{i}\bar{\nu}} (a_{\bar{i}}^\dag + b_{\bar{\nu}}^\dag) \nonumber \\
	i \dot{b}_{s,\bar{i}\bar{\nu}} &=& \Delta_s b_{s,\bar{i}\bar{\nu}} + \Omega b_{e,\bar{i}\bar{\nu}} \nonumber \\
	i \dot{a}_{\bar{i}} &=& \Delta_{\bar{i}} a_{\bar{i}} + g \sum_\nu b^\dag_{g,\bar{i}\nu} b_{e,\bar{i}\nu} \nonumber \\
	i \dot{b}_{\bar{\nu}} &=& \Delta_{\bar{\nu}} b_{\bar{\nu}} + g \sum_i b^\dag_{g,i\bar{\nu}} b_{e,i\bar{\nu}},
\eeqa
where the summation over $\nu$ and $i$ in the last two equations is emphasized. Setting $\dot{b}_{e,\bar{i}\bar{\nu}} = 0$ and substituting to the remaining equations yields
\beqa
	i \dot{b}_{g,\bar{i}\bar{\nu}} &=& \left(-\frac{g^2}{\Delta_e}\right) \left( a_{\bar{i}}^\dag + b_{\bar{\nu}}^\dag  \right) \left( a_{\bar{i}} + b_{\bar{\nu}}  \right) b_{g,\bar{i}\bar{\nu}} + \left(-\frac{g\Omega}{\Delta_e}\right) \left( a_{\bar{i}}^\dag + b_{\bar{\nu}}^\dag  \right) b_{s,\bar{i}\bar{\nu}}  \nonumber \\
	i \dot{b}_{s,\bar{i}\bar{\nu}} &=& \Delta_s b_{s,\bar{i}\bar{\nu}} + \left(-\frac{\Omega^2}{\Delta_e}\right) b_{s,\bar{i}\bar{\nu}} + \left(-\frac{g\Omega}{\Delta_e}\right) b_{g,\bar{i}\bar{\nu}} \left( a_{\bar{i}} + b_{\bar{\nu}}  \right) \nonumber \\
	i \dot{a}_{\bar{i}} &=& \Delta_{\bar{i}} a_{\bar{i}} + \sum_\nu \left(-\frac{g^2}{\Delta_e}\right) b_{g,\bar{i}\nu}^\dag b_{g,\bar{i}\nu} \left( a_{\bar{i}} + b_{\nu}  \right) + \left(-\frac{g\Omega}{\Delta_e}\right) b^\dag_{g,\bar{i}\nu} b_{s,\bar{i}\nu} \nonumber \\
	i \dot{b}_{\bar{\nu}} &=& \Delta_{\bar{\nu}} b_{\bar{\nu}} + \sum_i \left(-\frac{g^2}{\Delta_e}\right) b_{g,i \bar{\nu}}^\dag b_{g,i \bar{\nu}} \left( a_{i} + b_{\bar{\nu}}  \right) + \left(-\frac{g\Omega}{\Delta_e}\right) b^\dag_{g,i \bar{\nu}} b_{s,i \bar{\nu}}.
\eeqa
Switching back to the Pauli matrices representation, now in the $\{\ket{s},\ket{g}\}$ basis, the effective Hamiltonian reads
\beq
	H = \Delta_i a_i^\dag a_i + \Delta_\nu b_\nu^\dag b_\nu + \frac{\tilde{\omega}_{a,i\nu}}{2} \sigma^z_{i\nu} + \tilde{g}\left( \sigma^+_{i\nu} (a_i + b_\nu) + {\rm h.c.} \right)+F_{i \nu},	
	\label{eq:H eff RWA 2D app}
\eeq
where
\beqa
	\tilde{\omega}_{a,\bar{i} \bar{\nu}} &=& \Delta_s - \frac{\Omega^2 - g^2 \left( a_{\bar{i}}^\dag + b_{\bar \nu}^\dag  \right) \left( a_{\bar i} + b_{\bar \nu}  \right) }{\Delta_e} \nonumber \\
	\tilde{g} &=& -\frac{g \Omega}{\Delta_e} \nonumber \\
	F_{\bar{i} \bar{\nu}} &=& \frac{1}{2} \left( \Delta_s - \frac{\Omega^2 + g^2 \left( a_{\bar i}^\dag + b_{\bar \nu}^\dag  \right) \left( a_{\bar i} + b_{\bar \nu}  \right) }{\Delta_e} \right).
\eeqa
The corresponding equations of motion for the operators read
\begin{subequations}
\label{eq:op EoMs}
\begin{align}
	i \dot{a}_{\bar{i}} &= \left( \Delta_{\rm ph}^{\bar{i}} - i \kappa +  \lambda \sigma^z_{\bar{i} \nu} \right) a_{\bar{i}} + \tilde{g} \sigma^-_{\bar{i} \nu} + \lambda b_{\nu} \left( \sigma^z_{\bar{i} \nu} - \mathds{1} \right) \label{eq:op EoMs a} \\
	i \dot{b}_{\bar{\nu}} &= \left( \Delta_{\rm ph}^{\bar{\nu}} - i \kappa + \lambda \sigma^z_{i \bar{\nu}} \right) b_{\bar{\nu}} + \tilde{g} \sigma^-_{i \bar{\nu}} + \lambda a_{i} \left( \sigma^z_{i \bar{\nu}} - \mathds{1} \right) \label{eq:op EoMs b} \\
	i \dot{\sigma}^-_{\bar{i} \bar{\nu}} &= \left[ \Delta_{\rm at} -i\frac{\gamma}{2} + 2 \lambda \left( a^\dag_{\bar{i}} + b^\dag_{\bar{\nu}} \right) \left( a_{\bar{i}} + b_{\bar{\nu}} \right) \right] \sigma^-_{\bar{i} \bar{\nu}} - \tilde{g} \left( a_{\bar{i}} + b_{\bar{\nu}} \right) \sigma^z_{\bar{i} \bar{\nu}} \label{eq:op EoMs sm} \\	
	i \dot{\sigma}^z_{\bar{i} \bar{\nu}} &= 2\tilde{g} \left[ \sigma^+_{\bar{i} \bar{\nu}} \left( a_{\bar{i}} + b_{\bar{\nu}} \right) - {\rm h.c.} \right] - i \gamma \left( \sigma^z_{\bar{i} \bar{\nu}} + \mathds{1} \right), \label{eq:op EoMs sz}
\end{align}
\end{subequations}
where
\beqa
	\Delta_{\rm at} &=& \Delta_s - \frac{\Omega^2}{\Delta_e} \nonumber \\
	\Delta_{\rm ph}^l &=& \Delta_l - \frac{g^2}{2 \Delta_e} \nonumber \\
	\lambda &=& -\frac{g^2}{2 \Delta_e}
\eeqa
and we have introduced the cavity and spin decays $\kappa, \gamma$ which we take to be the same for all cavity modes and all spins respectively.

\section{General stability matrix}
\label{app:General Stability}

The matrix $M$ used in the stability study, equation (\ref{eq:stability}), can be simply obtained from the MF equations of motion (\ref{eq:EoMs MF pump real}) and reads
\beq
  M = \begin{bmatrix}
       -\kappa & \Delta_{\rm ph} + \lambda \bar{w} & 0 & -\frac{\tilde{g}}{2} & \lambda \bar{\alpha}_I \\
       -\Delta_{\rm ph} - \lambda \bar{w} & -\kappa & -\frac{\tilde{g}}{2} & 0 & -\lambda \bar{\alpha}_R \\
       -4\lambda \bar{\alpha}_R \bar{s}_y & -2 \tilde{g} \bar{w} -4\lambda \bar{\alpha}_I \bar{s}_y & 0 & -\left( \Delta_{\rm at} + 2 \lambda |\bar{\alpha}|^2 \right) & -2 \tilde{g} \bar{\alpha}_I \\
       -2 \tilde{g} \bar{w} +4\lambda \bar{\alpha}_R \bar{s}_x & 4\lambda \bar{\alpha}_I \bar{s}_x & \left( \Delta_{\rm at} + 2 \lambda |\bar{\alpha}|^2 \right) & 0 & -2 \tilde{g} \bar{\alpha}_R \\
       2 \tilde{g} \bar{s}_y & 2 \tilde{g} \bar{s}_x & 2 \tilde{g} \bar{\alpha}_I & 2 \tilde{g} \bar{\alpha}_R & 0        
      \end{bmatrix},
\eeq
where $|\bar{\alpha}|^2 = \bar{\alpha}_R^2 + \bar{\alpha}_I^2$ and $\bar{v}$ are the steady state solutions, $\bar{v} \in \{ \bar{\alpha}_R, \bar{\alpha}_I, \bar{s}_x, \bar{s}_y, \bar{w} \}$.

\section{Large $\lambda$ expansion}
\label{app:lambda exp.}

Here we seek a perturbative solution of the algebraic steady state equations (\ref{eq:spin conservation}) and (\ref{eq:SS lneq0 D0}) in the large $\lambda$ limit. This can be done using a perturbative ansatz for the variables of the form
\beq
	v = \sum_{n=0}^\infty \lambda^{-n} v^{(n)} = v^{(0)} + \lambda^{-1} v^{(1)} + \lambda^{-2} v^{(2)} + ...
\eeq
The set of equations of order $\lambda^1$ read
\beqa
	\alpha_I^{(0)} w^{(0)} &=& 0 \nonumber \\
	\alpha_R^{(0)} w^{(0)} &=& 0 \nonumber \\
	\left( \left. \alpha_I^{(0)}\right.^2 + \left.\alpha_I^{(0)}\right.^2 \right) s_y^{(0)} &=& 0 \nonumber \\
	\left( \left. \alpha_I^{(0)}\right.^2 + \left.\alpha_I^{(0)}\right.^2 \right) s_x^{(0)} &=& 0,
	\label{eq:exp order -1}
\eeqa
which yield the solution for $\alpha_R^{(0)} = \alpha_I^{(0)} = 0$. Using this result, order $\lambda^0$ equations simplify to
\begin{subequations}
\label{eq:exp order 0}
\begin{align}
	\alpha_R^{(1)} &= - \frac{\tilde{g}s_x^{(0)}+2\eta_R}{2 w^{(0)}} \label{eq:exp order 0 aR} \\
	\alpha_I^{(1)} &= \frac{\tilde{g}s_y^{(0)}-2\eta_I}{2 w^{(0)}} \label{eq:exp order 0 aI} \\	
	\left. s_x^{(0)}\right.^2 + \left. s_y^{(0)}\right.^2 + \left. w^{(0)}\right.^2 &= N^2. \label{eq:exp order 0 cons}
\end{align}
\end{subequations}
In order to proceed, we realize that the equation (\ref{eq:SS lneq0 D0 w}) at order $\lambda^{-1}$ reads
\beqa
	\alpha_R^{(1)} s_y^{(0)} = - \alpha_I^{(1)} s_x^{(0)}.
\eeqa
Substituting for $\alpha_R^{(1)},\alpha_I^{(1)}$ from (\ref{eq:exp order 0 aR},\ref{eq:exp order 0 aI}), it can be cast to the form
\beq
	-\frac{s_y^{(0)}}{s_x^{(0)}} = \frac{\eta_I}{\eta_R}.
\eeq
In order to simplify the treatment further, we put $\eta_I=0$ which implies $s_y^{(0)}=0$ and by the sake of (\ref{eq:exp order 0 aI}) $\alpha_I^{(1)}=0$. We will use these solutions in what follows. 

Order $\lambda^{-1}$ equations, after the substitutions of the solutions $\alpha_R^{(0)} = \alpha_I^{(0)}=0$ read
\begin{subequations}
\label{eq:exp order 1}
\begin{align}
	0 &= -\kappa \alpha_R^{(1)} + \left( \Delta_{\rm ph} + w^{(1)}\right) \alpha_I^{(1)} - \frac{\tilde{g}}{2} s_y^{(1)} + \alpha_I^{(2)} w^{(0)} \\
	0 &= -\left( \Delta_{\rm ph} + w^{(1)}\right) \alpha_R^{(1)} -\kappa \alpha_I^{(1)} - \frac{\tilde{g}}{2} s_x^{(1)} - \alpha_R^{(2)} w^{(0)} \\
	0 &= \tilde{g} \alpha_I^{(1)} w^{(0)} + \left|\alpha^{(1)} \right|^2 s_y^{(0)} \label{eq:exp order 1 sy} \\
	0 &= \tilde{g} \alpha_R^{(1)} w^{(0)} - \left|\alpha^{(1)} \right|^2 s_x^{(0)} \label{eq:exp order 1 sx} \\
	0 &= w^{(0)} w^{(1)} + s_x^{(0)} s_x^{(1)} + s_y^{(0)} s_y^{(1)}.
\end{align}
\end{subequations}  
When substituting the solutions for $s_y^{(0)} = \alpha_I^{(1)}=0$, (\ref{eq:exp order 1 sy}) is trivial and the equations (\ref{eq:exp order 1}) simplify to
\begin{subequations}
\label{eq:exp order 1 simp}
\begin{align}
	0 &= -\kappa \alpha_R^{(1)} + - \frac{\tilde{g}}{2} s_y^{(1)} + \alpha_I^{(2)} w^{(0)} \\	
	0 &= -\left( \Delta_{\rm ph} + w^{(1)}\right) \alpha_R^{(1)} - \frac{\tilde{g}}{2} s_x^{(1)} - \alpha_R^{(2)} w^{(0)} \\	
	0 &= \tilde{g} \alpha_R^{(1)} w^{(0)} - \left.\alpha_R^{(1)} \right.^2 s_x^{(0)} \label{eq:exp order 1 simp sx} \\
	0 &= w^{(0)} w^{(1)} + s_x^{(0)} s_x^{(1)}.
\end{align}
\end{subequations}
Substituting the solution for $\alpha_R^{(1)}$ from (\ref{eq:exp order 0 aR}) into (\ref{eq:exp order 1 simp sx}) yields
\beq
	\left( \tilde{g} s_x^{(0)} + 2 \eta_R \right) \left[ 2 \tilde{g} \left. w^{(0)}\right.^2 + \left( \tilde{g} s_x^{(0)} + 2 \eta_R \right) s_x^{(0)} \right] = 0
\eeq
which has the solutions
\begin{subequations}
\label{eq:sol sx order 1}
\begin{align}
	s_x^{(0)} &= -\frac{2 \eta}{\tilde{g}} \label{eq:sol sx order 1a} \\
	s_{x,\pm}^{(0)} &= \frac{\eta}{\tilde{g}} \left(1 \pm \sqrt{1+\frac{2 \tilde{g}^2 N^2}{\eta^2}}\right), \label{eq:sol sx order 1b}
\end{align}
\end{subequations}
where $\eta=\eta_R$ and we have used the spin conservation $\left. w^{(0)} \right.^2 + \left. s_x^{(0)} \right.^2 = N^2$. Since $|s_x^{(0)}| \leq N$, the solution (\ref{eq:sol sx order 1a}) is valid for $\tilde{g} \geq g_1^* = 2 \eta/N$. Next, we remark, that the function $s_{x,+}^{(0)}(\tilde{g})$ is monotonously decreasing with the limit
\beq
	\lim_{\tilde{g} \to \infty} s_{x,+}^{(0)}(\tilde{g}) = \sqrt{2} N > N,
\eeq
i.e. is unphysical. We are thus left with the solution $s_{x,-}^{(0)}(\tilde{g})$ which is also monotonously decreasing function with the asymptotes
\beqa
	\lim_{\tilde{g} \to 0} s_{x,-}^{(0)}(\tilde{g}) &=& 0 \nonumber \\
	\lim_{\tilde{g} \to \infty} s_{x,-}^{(0)}(\tilde{g}) &=& -\sqrt{2} N.
\eeqa
It is thus clear that the solution $s_{x,-}^{(0)}$ is valid for $g \in [0,g^*]$, where $g^*$ is some critical value for which $s_{x,-}^{(0)}$ reaches the physically allowed maximum $\left| s_{x,-}^{(0)}\right| = N$. It is easy to find, that $g^* = g_1^*$. This completes the leading order solutions and yields the expressions (\ref{eq:sx0 1}),(\ref{eq:sx0 2}).

The equations (\ref{eq:exp order -1}),(\ref{eq:exp order 0}),(\ref{eq:exp order 1}) yield a closed set for spin variables up to the order $\lambda^{-1}$ and for $\alpha$ up to $\lambda^{-2}$. It is straightforward to find the solutions up to the respective order explicitly, giving however rather lengthy algebraic expressions. These can be simplified in specific situations. For example, as we discuss in the main text, when looking for solutions in the vicinity of the transition point $g_1^* = 2\eta/N$, the leading order contribution to $|\alpha|^2$ is of order $\lambda^{-2}$, namely
\beq
	\left| \alpha \right|^2 = \frac{1}{\lambda^2} \left. \alpha_R^{(1)} \right.^2 + O(\lambda^{-3}),
\eeq 
which gives the relation (\ref{eq:a2}). Similarly, the leading contribution to $w$ is of order $\lambda^0$ yielding the expressions (\ref{eq:w0}).

\end{appendices}

\end{document}